\documentclass[twocolumn,twocolappendix]{aastex631} 
\usepackage{graphicx} 
\usepackage[caption=false]{subfig}
\usepackage{float}

\usepackage{amsmath}

\begin{document}

\newcommand{\excessdiff}{$0.00 \pm 0.05$}
\newcommand{\sdt}[1]{{\color[rgb]{0.9,0.0,0.0} {\textbf{#1}}}}

\title{Searching for Bumps in the Cosmological Road: Do Type Ia Supernovae with Early Excesses Have Biased Hubble Residuals?}

\author[0000-0002-8559-0788]{Christine Ye}
\affiliation{Stanford University, 450 Jane Stanford Way, Stanford, CA 94305, USA}

\author[0000-0002-6230-0151]{David~O.~Jones}
\affiliation{Gemini Observatory, NSF's NOIRLab, 670 N Aohoku Pl, Hilo, HI, 96720}
\affiliation{Institute for Astronomy, University of Hawai‘i, 640 N.\ Aohoku Pl., Hilo, HI 96720, USA}

\author[0000-0003-3953-9532]{Willem~B.~Hoogendam}
\altaffiliation{NSF Graduate Research Fellow}
\affiliation{Institute for Astronomy, University of Hawai‘i, 2680 Woodlawn Dr., Honolulu, HI 96822, USA}

\author[0000-0003-4631-1149]{Benjamin~J.~Shappee}
\affiliation{Institute for Astronomy, University of Hawai‘i, 2680 Woodlawn Dr., Honolulu, HI 96822, USA}

\author[0000-0002-2376-6979]{Suhail Dhawan}
\affiliation{Institute of Astronomy and Kavli Institute for Cosmology, University of Cambridge, Madingley Road, Cambridge CB3 0HA, UK}

\author[0000-0002-0869-8760]{Sammy~N.~Sharief}
\affiliation{Department of Astronomy, University of Illinois at Urbana Champaign, 1002 W. Green St., IL 61801, USA}

\date{\today}

\begin{abstract}
Flux excesses in the early time light curves of Type Ia supernovae (SNe\,Ia) are predicted by multiple theoretical models and have been observed in a number of nearby SNe\,Ia over the last decade. However, the astrophysical processes that cause these excesses 
may affect their use as standardizable candles for cosmological parameter measurements.
In this paper, we perform a systematic search for early-time excesses in SNe\,Ia observed by the Zwicky Transient Facility (ZTF) to study whether SNe\,Ia with these excesses yield systematically different Hubble residuals. We analyze two compilations of ZTF SN\,Ia light curves from its first year of operations: 127 high-cadence light curves from \citet{Yao19} and 305 light curves from the ZTF cosmology data release of \citet{Dhawan22}. We detect significant early-time excesses for 17 SNe\,Ia in these samples and find that the excesses have an average $g-r$ color of $0.06\pm0.09$~mag; we do not find a clear preference for blue excesses as predicted by several models. Using the SALT3 model, we measure Hubble residuals for these two samples and find that excess-having SNe\,Ia may have 
lower Hubble residuals (HR) after correcting for shape, color, and host-galaxy mass, at $\sim$2-3$\sigma$ significance; our baseline result is $\Delta HR = -0.056 \pm 0.026$~mag ($2.2 \sigma$).  
We compare the host-galaxy masses of excess-having and no-excess SNe\,Ia and find they are consistent, though at marginal significance excess-having SNe\,Ia may prefer lower-mass hosts.
Additional discoveries of early excess SNe\,Ia will be a powerful way to understand potential biases in SN\,Ia cosmology and probe the physics of SN\,Ia progenitors.
\end{abstract}

\section{Introduction} \label{intro}
Type Ia supernovae (SNe\,Ia), the thermonuclear explosions of white dwarfs \citep{Hoyle60}, are highly standardizable distance indicators, making them powerful tools for precision cosmology \citep{Riess98,Perlmutter99}. However, while it is generally understood that SNe\,Ia occur from white dwarfs with binary companions, there is still uncertainty around how to explain the diverse range of observable SN\,Ia properties with different progenitor and explosion models (for reviews, see \citealt{Maoz14,Maeda16,Livio18,Jha19}). Importantly, if observational details (i.e., intrinsic brightness, color/shape calibration, etc.) vary as a function of galaxy properties or cosmic time, this could introduce systematic changes in their inferred distance measurements that have consequences for the accuracy of SN\,Ia-based cosmology experiments, especially if the population is changing with redshift  \citep[e.g.,][]{Childress14,Jones18,Rose19,Rigault20,Brout21,Nicolas21,Kelsey22,Thorp22,Kelsey23,Wiseman23}.

In recent years, measurements of cosmological parameters such as the Hubble Constant (H$_0$) and the dark energy equation of state parameter ($w$) have become increasingly precise \citep{Betoule14,Riess16,Burns18,Scolnic18,Freedman19,Jones19,Brout22,Riess22,Dhawan23,Garnavich23,Rubin23,Uddin23}. 
As these measurements have improved, indications of a $>$5$\sigma$ tension in H$_0$ between late-universe \citep{Riess22} and early-universe \citep{Planck18} experiments has appeared. This discrepancy potentially implies the existence of new physics (see \citealp{DiValentino21} for a review) and motivates the need for even more robust measurements in the coming years. For example, systematic uncertainties in SN\,Ia distances --- though they are highly unlikely to be of a size that resolves the H$_0$ tension \citep{Dhawan20} --- could stem from coincidental or systematic mismatches between SN\,Ia properties in different rungs of the H$_0$ distance ladder.  Similarly, changes in SN\,Ia properties across cosmic time have the potential to affect $w$ \citep[e.g.,][]{Pan22}.

As an example of such a systematic on $w$, an overluminous subpopulation of SNe\,Ia would produce relative underestimates of distances. While such SNe\,Ia may be rare in the local universe \citep{Desai23}, they may be more common at higher redshifts as cosmic metallicity decreases \citep[e.g.,][]{Dominguez01}, thereby affecting measurements of $w$.  The discovery of the (still poorly understood) dependence of SN\,Ia distance measurements on the masses of their host galaxies \citep{Kelly10,Lampeitl10,Sullivan10} is one previous systematic error in SN\,Ia distances that affected both $w$ and H$_0$ measurements \citep{Sullivan11,Riess16}.

Such systematics are not well constrained from a theoretical perspective; models for SNe\,Ia vary in the progenitor system (e.g., whether the companion is a white dwarf or a nondegenerate star; \citealp{Whelan73,Iben84,Webbink84,Livne90,Nomoto97}), and/or the explosion mechanism \citep[e.g.,][]{Whelan73,Nomoto80,Iben84,Woosley94,Hoeflich96,Piersanti03,vanKerkwijk10,Diamond18}. But while most physical models of SNe\,Ia generate indistinguishable predictions near maximum light, some disagree in their predictions for very early times post explosion (the first hours to days) and the very late time evolution \citep[e.g.,][]{Ashall19,Kumar23}. 

For example, in the single-degenerate case, the SN\,Ia ejecta may collide with the companion star, which becomes shock-heated and results in a blue/UV excess in the first 5 days after explosion that is observable in $\sim 10-20\%$ of events \citep{Marietta00,Kasen10,Burke22a,Burke22b,Deckers22}.  However, \citet{Hoogendam23} find that companion interaction is unlikely to explain the diversity of observed excesses; alternatively, models with circumstellar material, double-detonation mechanisms (an outer shell detonation that triggers a detonation in the core; \citealp{Fink07}), or radioactive matter in the outer ejecta \citep[e.g.,][]{Jiang18} also predict early excesses, although often with distinct color/spectral evolution \citep{Piro12,Piro16,Contreras18,Maeda18,Polin19,Magee20a,Magee20b}.

Prior to $\sim$2010, few SNe\,Ia could be observed with sufficient cadence or depth at early times to observe such signatures. However, over the last decade, there have been numerous reports of SNe\,Ia with early-time behavior that is inconsistent with a simple power-law rise. Some notable examples from the literature are SN~2012cg \citep[but also see \citealp{Shappee18}]{Marion16}, SN~2012fr \citep{Contreras18}, SN~2014J \citep{Goobar15,Siverd15}, SN~2016jhr \citep{Jiang17}, SN~2017cbv \citep{Hosseinzadeh17}, SN~2018oh \citep{Dimitriadis19,Li19,Shappee19}, SN~2021aefx \citep{Ashall22,Hosseinzadeh22} and, most recently, SN~2023bee \citep{Hosseinzadeh23,Wang23}, all of which exhibit a monotonic ``excess" above the power-law expectation, before a steeper power-law rise in luminosity. Reported excesses occur in normal, overluminous, and underluminous SNe, at UV, blue, and/or red wavelengths \citep[e.g.,]{Stritzinger18}, and with varied shapes \citep[e.g.,][]{Fausnaugh23}.

Other early-excess SNe\,Ia exhibit non-monotonic ``bumps," declining slightly before the power-law rise, such as SN~iPTF14atg \citep{Kroner16}, SN~2019yvq \citep{Miller20}, SN~2020hvf \citep{Jiang21}, SN~2021zny \citep{Dimitriadis23}, and SN~2022ilv \citep{Srivastav23}. Presently, only 2002es-like and 2003fg-like SNe\,Ia have observed bumps, and while the ultraviolet colors show significant differences between bump and other SNe\,Ia, the optical colors are indistinguishable between the two categories, suggesting that differing mechanisms produce ``excess'' and ``bump'' SNe\,Ia \citep{Hoogendam23}.

Given the potential for different physics in these early-excess explosions, possibly leading to systematic biases, it is important to check whether their distance measurements can reliably be used for cosmology.  In this paper, we examine whether SNe\,Ia with early excesses have systematic offsets in their distance measurements compared to non-excess SNe\,Ia, by systematically identifying SNe\,Ia with early excesses in Zwicky Transient Facility light curves \citep[ZTF;][]{Bellm19}.  We also search for population differences between SNe\,Ia with and without early excesses.

The text is organized as follows: in Section \ref{data} we describe the light-curve fitting and sample selection procedures; in Section \ref{fitting} we detail criteria for excess and no-excess SNe\,Ia; and in Section \ref{results} we discuss and compare our results.  In Section \ref{sec:discussion} we discuss caveats and implications of our results.  Section \ref{conclusions} summarizes the results of the work and provides concluding remarks.

\section{Sample Selection} \label{data}
\subsection{ZTF Data}
The dataset used in this paper is taken from observations by the Zwicky Transient Facility (ZTF) in 2018 \citep{Bellm19,Masci19}.  We also investigate the $z < 0.05$ ZTF SNe\,Ia sample from 2019--2022, but exclude it from our analysis due to higher scatter in the Hubble residuals (future data releases will likely improve the fidelity of the light-curves for cosmological measurements) and a lack of early excess detections.

ZTF detects astronomical transients in an untargeted, high-cadence survey covering the $g, r,$ and $i$ bands. At the time of the 2018 observations, ZTF's Mid-Scale Innovations Program (MSIP) survey was covering approximately 13,000 deg$^2$ per night in the $gr$ bands at a cadence of three days (the ZTF phase II survey, which commenced in late 2020, has a two-day cadence).  Additionally, ZTF conducted an $i$-band survey across $\sim$8000 deg$^2$ with a four-day cadence and a one-day cadence survey in $gr$ across $\sim$1700 deg$^2$.  Additional information about the ZTF surveys is given by \citet{Bellm19}. 

We use data published by \citet{Yao19} and \citet{Dhawan22}, which is from a combination of the ZTF surveys described above.  The \citet{Yao19} sample specifically selects SNe\,Ia with high-S/N, high-cadence early time light curves.  They use point spread function (PSF) fitting \citep{Masci19}, with the centroid fixed at the average SN\,Ia location, to produce light curves for 127 SNe\,Ia with detections earlier than 10 rest-frame days before maximum light, including 50 objects with detections earlier than 14 rest-frame days before maximum light.  In addition to the \citet{Yao19} analysis, the early time behavior of SNe\,Ia in this sample has also been investigated by \citet{Burke22a} and \citet{Deckers22}.  This sample has a median redshift of $z = 0.074$.   

The \citet{Dhawan22} analysis, from a total of 761 spectroscopically classified SNe\,Ia that were observed in 2018, re-reduces 305 SNe\,Ia with host-galaxy redshifts using a new photometric pipeline intended for cosmological analyses. The median redshift is $z = 0.057$.


Due to the poor sampling and reduced depth of $i$ band data, we consider only $g$ and $r$ light curves for identifying early excesses. 


\subsection{Light-curve Fitting for Cosmology}

We use the Python package \texttt{SNCosmo} to perform light-curve fitting with the \texttt{SALT3} spectral energy distribution model \citep{Kenworthy21,Taylor23}. \texttt{SALT3} improves upon previous generations of \texttt{SALT} (Spectral Adaptive Light-curve Template; \citealp{Guy05,Guy07}) models with an expanded training sample and better handling of uncertainties. We correct for Milky Way dust extinction using the \citet{Schlafly11} reddening maps. With redshifts from host galaxies (where available) or SN\,Ia spectra, \texttt{SALT3} fitting produces estimates for $x_0$, an overall amplitude parameter; $x_1$, the first principal component of variation, which correlates with shape; $c$, a color parameter; and $t_0$, the time of light-curve peak. $m_B$, the apparent $B$-band magnitude at peak, is directly estimated from \texttt{SALT3} $x_0$: $m_B = -2.5 \log_{10} x_0 + 10.635$~mag.

\subsection{Host-galaxy Mass Determination}

Due to the observed dependence of SN\,Ia standardized magnitudes on the host-galaxy stellar mass of a SN\,Ia \citep{Kelly10,Lampeitl10,Sullivan10}, SN\,Ia distance determinations must include host galaxy identifications and mass estimates.  Here, we match each SN\,Ia to its most likely host galaxy using the Galaxies HOsting Supernovae and other Transients (GHOST; \citealp{Gagliano21}) software.  GHOST applies a gradient ascent method to postage-stamp images of each SN\,Ia location to find the most likely host galaxy for the SN.

GHOST also provides Pan-STARRS catalog photometry \citep{Flewelling20} for these hosts, which we supplement with Sloan Digital Sky Survey (SDSS; \citealp{York00}) photometry when Pan-STARRS magnitudes are not available (Pan-STARRS images exist at every location with SDSS imaging, but we found that catalog photometry for a give source is sometimes unavailable).  In the rare cases when Pan-STARRS images exist, but Pan-STARRS and SDSS catalog magnitudes do not, we use elliptical aperture photometry within the isophotal radius to measure the galaxy's magnitudes from the Pan-STARRS images directly.

Galaxy masses were estimated using $g$ and $i$ band Kron magnitudes from the relation given by \citet{Taylor11}: 
\begin{equation}
    \log M_{*}/[M_\odot]  = 1.15 + 0.7(g - i) - 0.4(i - \mu(z))
\end{equation}
with $\mu(z)$ from \citet{Planck18} cosmology.  SNe\,Ia for which a host galaxy cannot be identified visually or with GHOST are assigned a host-galaxy mass $< 10$~dex.

\subsection{Distances and Cosmological Sample Selection} \label{sample}
For each SN, we estimate the distance modulus, $\mu$, from \texttt{SALT3} parameters using the Tripp equation \citep{Tripp98}:

\begin{multline}
    \mu = m_B + \alpha x_1 - \beta c -\mathcal{M}(z) + \gamma \mathcal{H}(M_\odot - 10),
    \label{eqn:tripp}
\end{multline} 

\noindent where $\alpha$ and $\beta$ are nuisance parameters that relate the $x_1$ and $c$ parameters to $\mu$, $\gamma$ is the size of the host-galaxy mass step, and $\mathcal{H}$ is the Heaviside step function.  $\mathcal{M}$ is the SN\,Ia absolute magnitude (degenerate with H$_0$). 

To remove any dependence of the Hubble residuals on redshift, including dependence on the cosmological model and distance-dependent selection effects \citep[e.g.,][]{Kessler17}, we fit $\mathcal{M}(z)$ as a piecewise function with independent values for $\mathcal{M}(0 < z \leq 0.033)$, $\mathcal{M} (0.033 < z \leq 0.067)$, and $\mathcal{M} (0.067 < z \leq 0.1)$.

The uncertainty on $\mu$, $\sigma_{\mu}$, is:

\begin{multline}
    \sigma_\mu^2 = \sigma_{int}^2 +  \sigma_{\mu,z}^2 + \sigma_{m_B}^2 + (\alpha \sigma_{x_1})^2 + (\beta \sigma_c)^2 \\
    + 2\alpha \beta \sigma_{c, x_1} + 2\alpha \sigma_{m_B, x_1} + 2\beta \sigma_{m_B, c}.
\end{multline}

\noindent We adopt $\sigma_{lens} = 0.055z$ following \citet{Jonsson10} and the covariances/errors $\sigma_{c, x_1}$, $\sigma_{m_B, x_1}$, $\sigma_{m_B, c}$, $\sigma_{x_1}$, $\sigma_c$ from the \texttt{SALT3} fits.  The intrinsic dispersion, $\sigma_{int}$, is the remaining scatter of SN\,Ia Hubble residuals after photometric and model uncertainties have been taken into account.  Corrections for the effect of local overdensities (peculiar velocities) are estimated from 2M$++$ \citep{Carrick15}, with uncertainties on the corrections of $250~{\rm km s^{-1}}$ \citep{Scolnic18}.

To standardize our sample, we impose the following selection requirements following previous SNe\,Ia cosmology analyses \citep[e.g.,][]{Brout22}.  To be included, SNe\,Ia must have:
\begin{enumerate}
    \item Normal or 1991T-like spectroscopic classifications.
    \item At least 3 data points between $-10$ and 10 days.
    \item Well-constrained shape: $\sigma_{x1} < 1$.
    \item Multiple filters (both $g$ and $r$) to constrain the color parameter and $\sigma_c < 0.3$.
    \item Well-constrained peak epoch: $\sigma_{t_0} < 1$.
    \item The $x_1$ parameter is consistent with a normal SN\,Ia: $-3 < x_1 < 3$.
    \item Color parameter consistent with a normal SN\,Ia that is not highly reddened: $-0.3 < c < 0.3$.
    \item Chauvenet's criterion: SNe\,Ia with less than a 50\% chance of belonging to the observed, normally-distributed SN\,Ia population.  This selection criterion is computed from the sample size and Hubble residual dispersion of our data and removes one SN\,Ia from each of the \citet{Yao19} and \citet{Dhawan22}  samples.
\end{enumerate} 
\noindent We also include a redshift cut of $0.01 < z < 0.1$, to ensure that peculiar velocities don't dominate the Hubble residual uncertainty at the low-redshift end, and to limit cosmological model sensitivity and distance biases at the high-redshift end.  

After these cuts, we are left with 194/305 SNe\,Ia from the \citet{Dhawan22} sample and 88/127 SNe\,Ia from the \citet{Yao19} sample. 
We simultaneously fit estimates of $\alpha, \beta, \gamma, \mathcal{M}(z)$, and $\sigma_{int}$ to both samples.  The results are presented in Table \ref{table:tripp}.  Although the intrinsic dispersion is higher than usual for a SN\,Ia sample (typically $\sim$0.1-0.12~mag), the total r.m.s.\ of the Hubble residuals is not; this is most likely an indication of under-estimated photometric errors.  We note that the \citet{Yao19} and \citet{Dhawan22} $\beta$ parameters are slightly inconsistent with each other, with the \citet{Yao19} value slightly closer to a typical SALT3 value \citep{Kenworthy21}.
This may be due to a color-dependent calibration error or selection effects, which differ substantially between the two samples and result in significantly different values for $\mathcal{M}(z)$; we discuss potential inconsistencies between the samples and possible implications of miscalibration on our results in Section \ref{sec:discussion}.

\begin{table*}[ht]
\centering
\begin{tabular}{ |c|c|c|c|c|c|c|} 
 \hline
 Data & $\alpha$ & $\beta$ & $\mathcal{M}$& $\gamma$ & $\sigma_{int}$ & rms \\
&&& (mag)& (mag)& (mag)& (mag)\\
 \hline
 \citet{Dhawan22} & $0.118 \pm 0.009$ & $2.54 \pm 0.08$ & $-19.07 \pm 0.02$, $-19.16 \pm 0.02$, $-19.21 \pm 0.02$ & $0.05 \pm 0.02$ & 0.144 & 0.152 \\
 \citet{Yao19} & $0.109 \pm 0.004$ & $2.773 \pm 0.004$ & $-19.25 \pm 0.05$, $-19.16 \pm 0.03$, $-19.18 \pm 0.02$ & $0.04 \pm 0.05$ & 0.149 & 0.153\\
 \hline
\end{tabular}
\caption{Nuisance parameters for SNe\,Ia in the two public samples analyzed in this work.  $\mathcal{M}$ is listed in order of increasing redshift range: $[0 < z \leq 0.033, 0.033 < z \leq 0.067, 0.067 < z \leq 0.1]$.}
\label{table:tripp}
\end{table*}

\section{Early Excess Fitting} \label{fitting}
\subsection{Models}
In the first few days after explosion, we would expect the flux of SNe\,Ia to follow a power-law rise of roughly $f \propto t^2$ under the assumption of a constant-temperature photosphere expanding at a constant rate \citep[the so-called expanding fireball model;][]{Riess99}. Indeed, studies have shown that this holds approximately true at least 10-15 days before maximum light \citep{Miller20}, although there are some notable exceptions \citep[e.g.,][]{Shappee16,Fausnaugh23, Hoogendam23}. At later times, closer to maximum light, the slope decreases. For SNe\,Ia with detected early excesses, multiple functional forms for light-curve fits have been recorded in the literature, including double power laws \citep[e.g.,][]{Shappee19}, single power laws with a Gaussian component \citep[e.g.,][]{Dimitriadis19}, and physically motivated models; for example, \citet{Hosseinzadeh23} fits light curves with a single-degenerate model that depends on parameters such as progenitor radius and shock velocity. 

After testing multiple fitting procedures on SNe\,Ia with known excesses, we find that a two-component (Gaussian and Power Law, hereafter the Gaussian$+$PL model) model offers sufficient flexibility in fitting all observed excesses, while also allowing for quantitative measures of the significance of the excess. This formulation also more directly corresponds to physical parameters in the system, such as the total amount of energy in the early excess, as compared to other analytic models.  We characterize early light curves for a given band $b$ as follows:
\begin{multline}
    f_b(t) = B_b + \mathcal{H}(t_{exp})A_b(t-t_{exp})^{\alpha_{b}} + C_b \mathcal{N}(\mu, \sigma)
\end{multline}
where $B_b$ is a constant offset, $t_{exp}$ refers to the time of first light from the explosion, $\mu$ and $\sigma$ are the location and standard deviation of the Gaussian component, $A_b$ is the power-law scale factor in a given band, $\alpha_b$ is a band-dependent power-law exponent, and $C_b$ is the Gaussian scale factor in a given band. $\mathcal{H}$ is the Heaviside step function, equal to zero before $t_{exp}$ and a constant afterwards.  In total, we have 9 parameters, which we fit for using least-squares minimization with constraints set in Table \ref{table:prior}.

\begin{table}[h!]
\centering
\caption{Constraints on Early Excess Fitting Parameters}
\begin{tabular}{ |c|c|} 
 \hline
 $B_r$ & [-100.0, 100.0] \\
 $B_g$ & [-100.0, 100.0] \\
 $A_r$ & [0.0, 500.0] \\
 $A_g$ & [0.0, 500.0] \\
 $t_{exp}$ & [$t_0 - 40$, $t_0 - 10$] \\
 $\alpha_r$ & [1.0, 3.0] \\
 $\alpha_g$ & [1.0, 3.0] \\
 $\mu$ & [$t_{exp}, t_{exp}$ + 5] \\
 $\sigma$ & [0.5, 4.0] \\
 $C_r$ & [0.0, 500.0] \\
 $C_g$ & [0.0, 500.0] \\
 \hline
\end{tabular}
\label{table:prior}
\end{table}

\noindent The parameter $t_0$ is the time of maximum light in the $B$ band, which we estimate using SALT3. We constrain $\mu$ to [$t_{exp}, t_{exp}$ + 5] because we are only searching for early excesses, and $\alpha$ to [1.0, 3.0] based on predicted and observed power-law light curve rises.

We also fit power-law-only (hereafter PL-only) models to the data to test whether a Gaussian component is truly necessary.  These models have seven parameters ($A_r, A_g, B_r, B_g, \alpha_r, \alpha_g, t_{exp}$).

For each SN\,Ia, we fit using the light-curve data between $t_0 - 40$ and $t_0 - N$, where $N$ ranges between 7 to 14 days. Allowing variation in $N$ between individual SNe\,Ia allows us to maximize the number of included data points while avoiding later epochs when the power-law flux model is no longer a valid approximation. The ideal value of $N$ is usually around $\sim 9-10$ days.  We visually inspect each identified early excess to ensure that the latest fitted epochs remain consistent with the best-fit power law.


\begin{figure*}
     \centering
     \subfloat{
         \centering
         \includegraphics[width=0.46\textwidth]{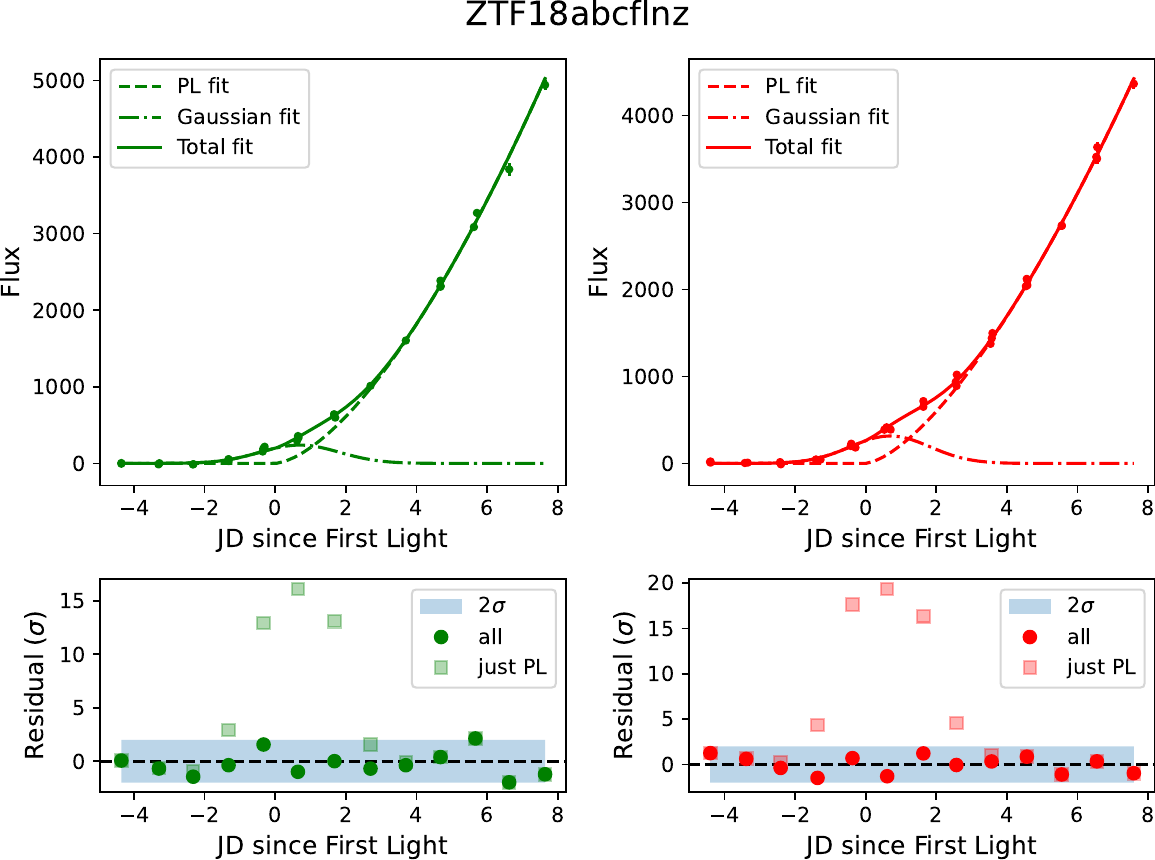}}
     \hfill
     \subfloat{
         \centering
         \includegraphics[width=0.46\textwidth]{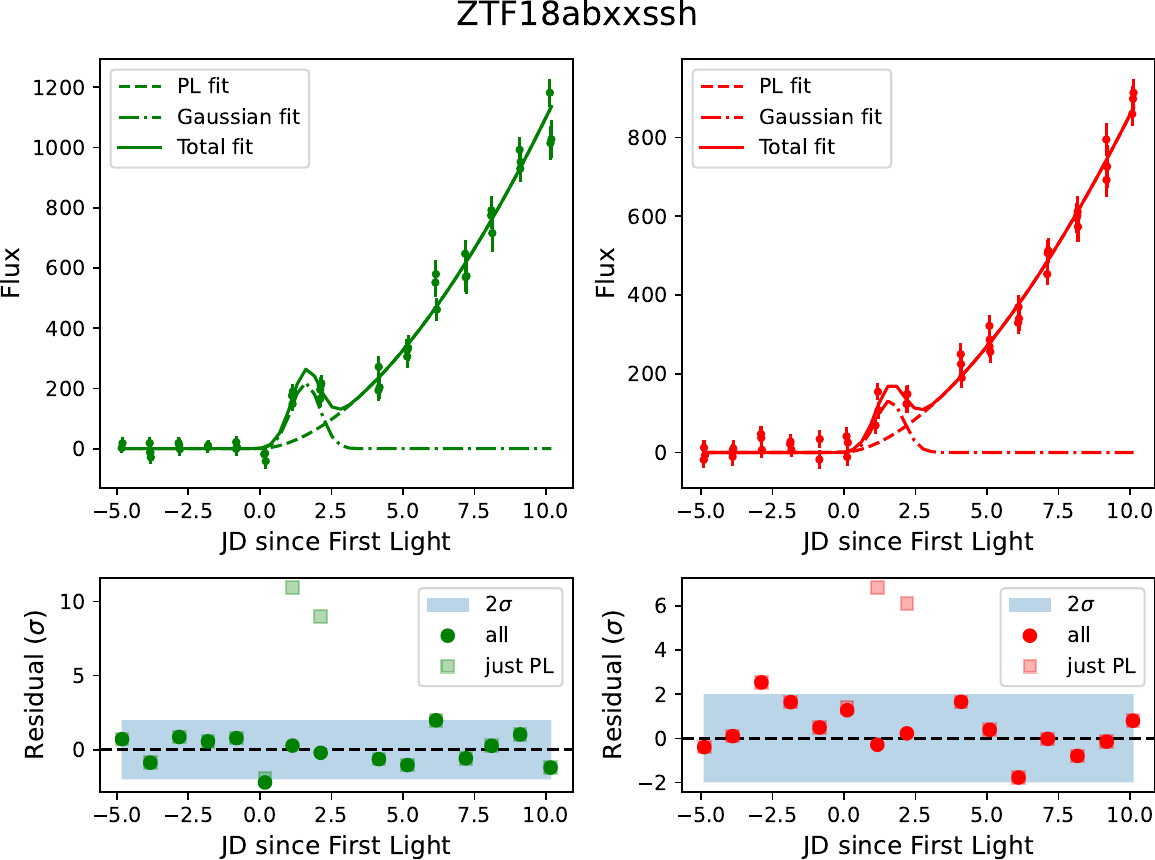}}
     
        \caption{Example gold-tier detections of bump early excesses. Panels display light curves in ZTF $g$ (green; left) and $r$ (red; right) bands, respectively. Points represent actual observations, while the lines represent fitted models: the power-law component (dashed), the Gaussian component (dash-dotted), and the total fit (solid). On epochs with multiple observations, the plotted residuals (bottom) are weighted averages. Fluxes are calculated with zero point set at 25.0 mag.}
        \label{fig:detection}
\end{figure*}

\subsection{Criteria for Detecting an Early Excess} \label{sec:tiers}
To systematically search for early-time excesses in SNe\,Ia, we require SNe\,Ia with potential early excesses to pass a number of statistical tests confirming their credibility. These specific criteria are given in the Appendix (\ref{sec:appendix}) and summarized below.  First, we check the quality and sampling of the light curve data, requiring that a substantial number of data points exist immediately pre- and post-explosion. In case excesses that appear statistically significant are due to poor light curve fits or underestimated flux errors, we also remove SNe\,Ia having fits with high reduced $\chi^2$ and/or individual extreme outliers. We measure the statistical significance of the early excess from the significance of the model fit's Gaussian component, and we visually inspect each light curve with a statistically significant excess as a final check on the quality of the data and the fit.

Moreover, models with more parameters will tend to fit any light curve better than models with fewer parameters, whether or not the additional parameters are actually necessary. Thus in order to calculate whether the Gaussian$+$PL model is truly favored over the PL-only model, we use the Bayesian Information Criterion (BIC), which penalizes models with more parameters:

\begin{figure*}
     \centering
     \subfloat{
         \centering
         \includegraphics[width=0.46\textwidth]{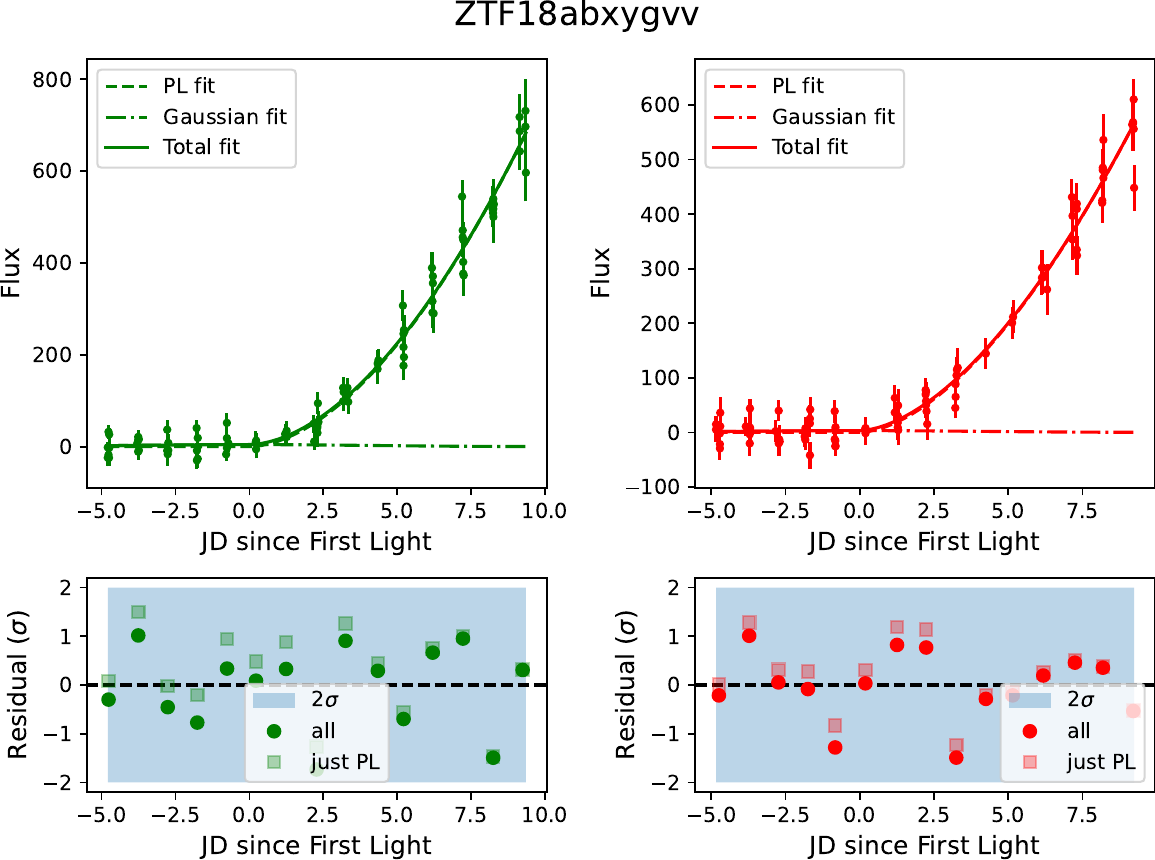}}
     \hfill
     \subfloat{
         \centering
         \includegraphics[width=0.46\textwidth]{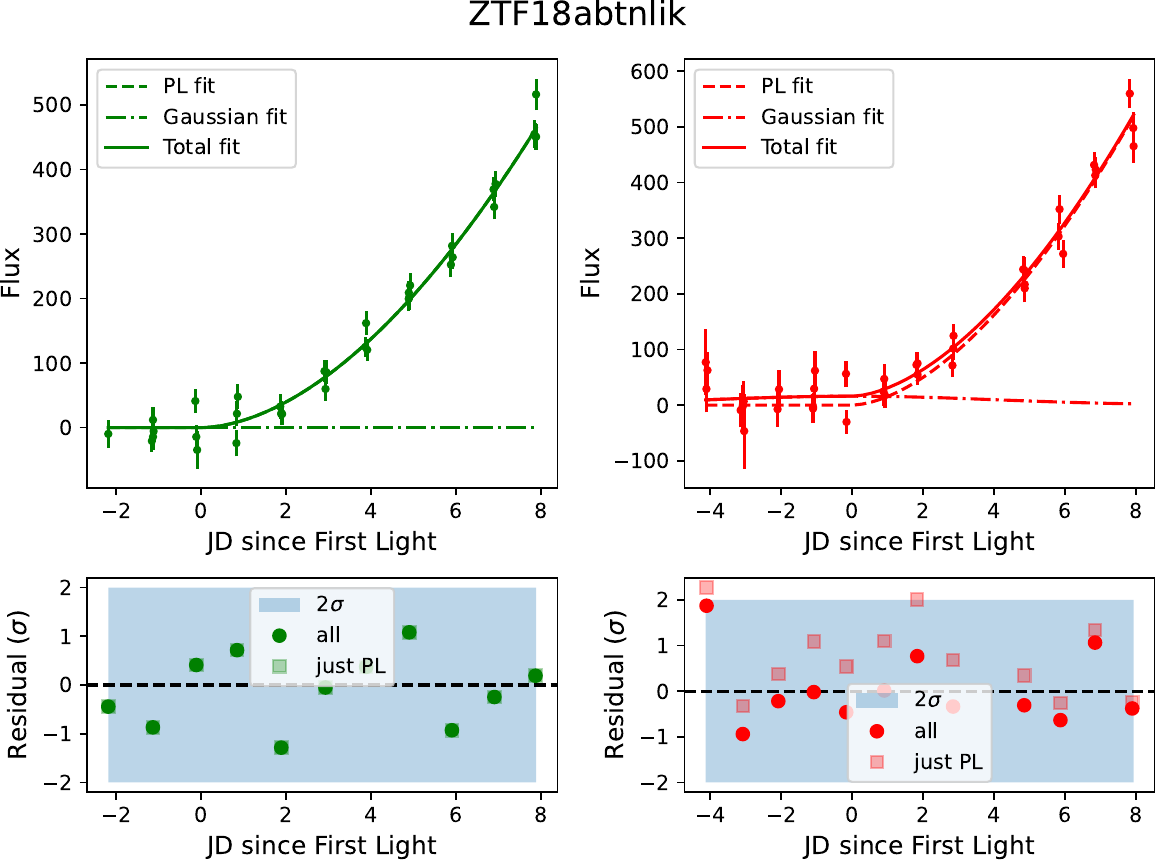}}
        \caption{Example gold-tier (left panel) and bronze-tier (right panel) non-detections of bump early excesses. Panels display light curves in ZTF $g$ (green; left) and $r$ (red; right) bands, respectively. 
        See the Figure \ref{fig:detection} caption for additional details.}
        \label{fig:nondetection}
\end{figure*}

\begin{equation}
    \text{BIC} = k \ln(n) - 2 \ln(\hat L),
\end{equation}
where $k$ is the number of parameters, $n$ is the number of data points, and $\hat L$ is the likelihood. The likelihood, which is assumed to be Gaussian, is given by: 

\begin{equation}
    \ln{\hat L} = \sum_i^n -\frac{(x_i - f_i(t))^2}{\sigma_{x,i}^2/2} + \ln(\frac{1}{\sqrt{2\pi \sigma{x_i}^2}}).
\end{equation}

\noindent Here, $x_i$ is the $i$th flux measurement, $f_i(t)$ is the model at the time of the $i$th flux, and $\sigma_{x,i}$ is the flux uncertainty. 

\begin{figure}
    \centering
    \includegraphics[width=3.5in]{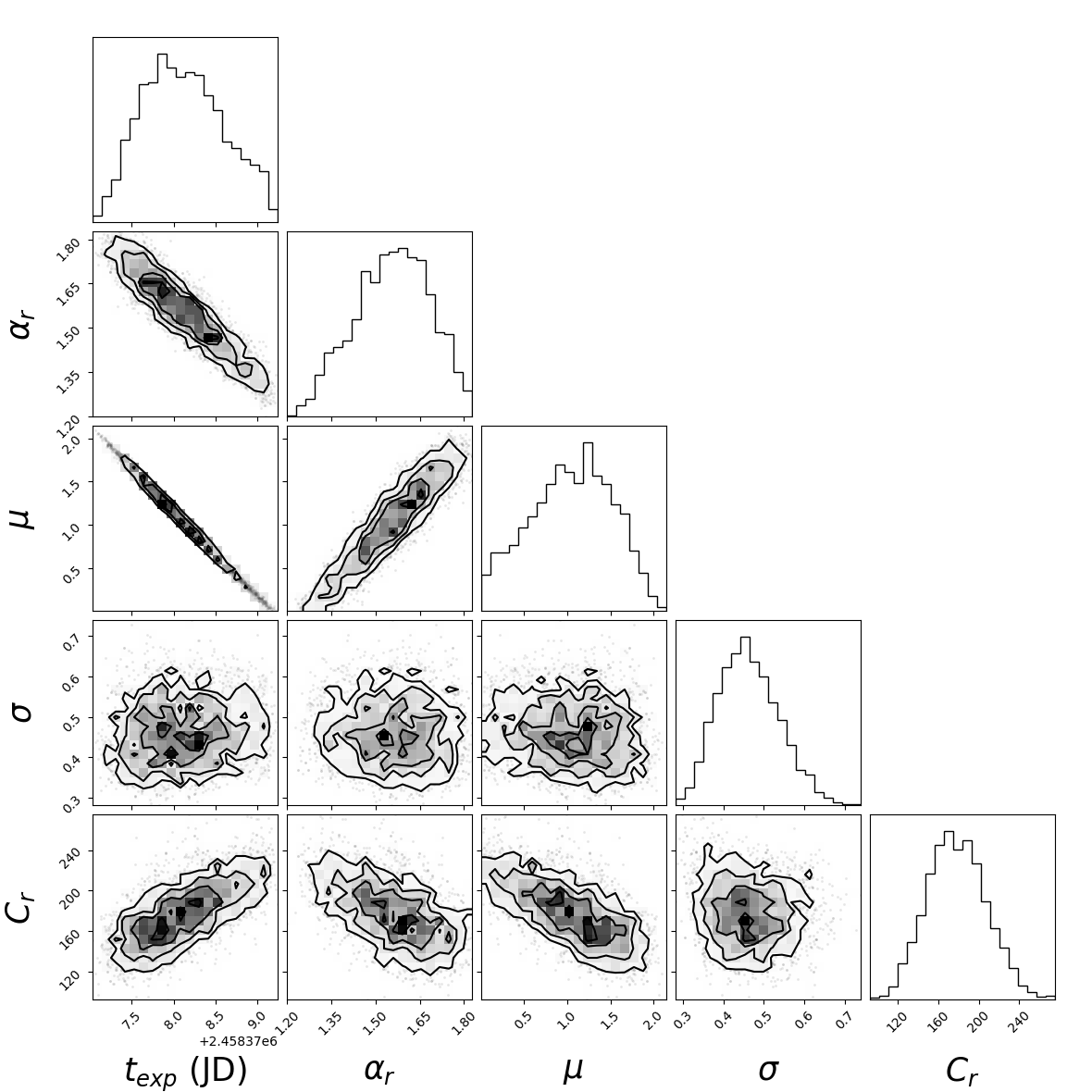}
    \caption{Corner plot showing a subset of parameters for ZTF18zzssh using PyMultiNest and {\tt corner} \citep{Buchner14,corner}. 
 From left to right: time of explosion ($t_{exp}$), $r$-band power-law slope ($\alpha_r$), Gaussian mean ($\mu$), Gaussian width ($\sigma$), and $r$-band Gaussian amplitude $C_r$.  Although the bump is detected with high significance, several correlations between these parameters are evident.}
    \label{fig:multinest}
\end{figure}

We also find that there is often a degeneracy between the power-law slope, $\alpha$, and the existence of an early excess.  An example of these degeneracies for one of our ``gold" SNe\,Ia, ZTF18abxxssh, is shown in Figure \ref{fig:multinest}.  Especially if the cadence is sparse, a lower value of $\alpha$ with a later explosion date and a non-zero excess component can often fit the data as well as a steeper power law, an earlier explosion date, and no excess component. This is particularly likely if $N$ is smaller, such that data closer to maximum light are being fit, increasing the chances that the light curve slope is no longer $\sim t^2$ at the later epochs\footnote{\citet{Olling15} find that a single power-law fit is a good approximation when the SN\,Ia flux is less than 40\% of its eventual maximum. Our chosen values of N are approximately in agreement with this cutoff but sometimes limited by data availability in days 0-5, as 40\% often corresponds to 5-6 days or earlier post-explosion.}. In these situations, we use the BIC to judge whether or not the additional ``bump" component is preferred. We also examine whether a potential early excess disappears when stricter bounds on $\alpha$ are placed, or some of the data nearer to maximum light is excluded (by increasing $N$).

Based on these considerations, we create two sets of criteria --- ``gold" and ``bronze" --- for defining different samples of SNe\,Ia with early excesses.  All identified gold and bronze tier SNe\,Ia from the \citet{Yao19} and \citet{Dhawan22} samples are given in Tables \ref{table:yaoexcess} and \ref{table:dhawanexcess}.  The exact criteria for inclusion in each of these samples is given in the Appendix (\ref{sec:appendix}).

The gold tier for non-detections is defined as SNe\,Ia for which we are confident that no excess exists at the flux limit required for inclusion in the gold sample.  However, in much of this analysis we also compare the ``gold" or ``bronze" excess-having sample to {\it all} SNe\,Ia for which an excess has not been detected (either due to the data's lack of sensitivity or an actual, significant non-detection of a ``bump").  
For this second sample, we note that if there is a Hubble residual difference between SNe\,Ia with excesses and those without (Section \ref{sec:hubbleres}), comparing to all SNe\,Ia without excess detections means that it would appear slightly smaller in this study than it would if we could identify 100\% of SNe\,Ia without early excesses.
However, because literature studies have estimated that just 10-20\% of SNe\,Ia have early excesses \citep[e.g.,][]{Burke22b}, the ``no excess" sample should not have a large contaminating fraction of excess-having SNe.  

We also considered the possibility that flux errors from forced photometry are underestimated, potentially resulting in the detection of ``significant" excesses that are not actually real. To account for this, we test the effect of multiplying the flux uncertainties by an amount such that the reduced $\chi^2$ of pre-explosion data, compared to the zero-flux expectation, is $\sim 1$. We find that this does not impact any claimed excesses from our sample.


Two high-cadence, low-scatter light curves with significant detections of early excesses are shown in Figure \ref{fig:detection} and are representative of our gold tier ``excess" sample. The remaining gold and bronze light curves are shown in the appendix.  Two example light curves with similar cadence and error bars but no ``bump" detection are shown in Figure \ref{fig:nondetection} and are representative of our gold tier ``no excess" sample. Due to the ZTF cadence, there are often multiple light-curve data points per night; assuming that the SN\,Ia flux does not vary drastically within a night, for plotting purposes we bin the residuals by epoch and take their weighted average. Moreover, for clarity, we only plot data points between $t_{exp} - 5$ and $t_0 - N$, but many light curves have long pre-explosion baselines that tightly constrain the epoch of explosion.

\begin{table*}[t]
\centering
\begin{tabular}{ |l|r|r|r|r|r|r|r|r|r| } 
  \hline
 
 ZTF ID & $z$ & $t_{0}$ & $m_B$ (mag) & $x_1$ & $c$ & $\Delta$ BIC & $f_{b, r}/f_{10, r}$ & $f_{b, g}/f_{10, g}$ &  $N$\\ 
 \hline
 \multicolumn{10}{|c|}{Gold Tier} \\
 \hline
 
aayjvve & 0.047 & 58292.3 & $17.655 \pm 0.005$ & $-0.18 \pm 0.03$ & $0.052 \pm 0.004$ & 27 & $0.07 \pm 0.02$ & $0.05 \pm 0.01$ & 8 \\
abcflnz & 0.027 & 58306.2 & $16.110 \pm 0.003$ & $-0.04 \pm 0.01$ & $-0.024 \pm 0.002$ & 31 & $0.03 \pm 0.01$ & $0.02 \pm 0.01$ & 9 \\
abucvbf & 0.055 & 58385.0 & $17.524 \pm 0.003$ & $-0.37 \pm 0.04$ & $-0.091 \pm 0.003$ & 36 & $0.05 \pm 0.01$ & $0.04 \pm 0.01$ & 8 \\
abxxssh & 0.078 & 58397.4 & $18.221 \pm 0.01$ & $1.51 \pm 0.10$ & $-0.035 \pm 0.007$ & 91 & $0.14 \pm 0.10$ & $0.20 \pm 0.10$ & 10 \\

 \hline
 \multicolumn{10}{|c|}{Bronze Tier} \\
 \hline
aasdted & 0.018 & 58265.5 & $15.838 \pm 0.002$ & $0.83 \pm 0.01$ & $0.166 \pm 0.002$ & 120 & $0.03 \pm 0.03$ & $0.02 \pm 0.03$ & 8 \\
aaslhxt & 0.055 & 58263.7 & $17.625 \pm 0.003$ & $0.36 \pm 0.02$ & $-0.116 \pm 0.002$ & 12 & $0.05 \pm 0.03$ & $0.05 \pm 0.03$ & 8 \\
aaxsioa & 0.032 & 58286.2 & $16.884 \pm 0.002$ & $-1.61 \pm 0.01$ & $0.063 \pm 0.002$ & $-5$ & $0.03 \pm 0.08$ & $0.00 \pm 0.01$ & 7 \\
aazsabq & 0.060 & 58294.0 & $18.233 \pm 0.004$ & $-1.46 \pm 0.03$ & $0.033 \pm 0.004$ & $-10$ & $0.04 \pm 0.03$ & $0.01 \pm 0.01$ & 8 \\
abauprj & 0.024 & 58302.6 & $15.526 \pm 0.002$ & $1.37 \pm 0.02$ & $-0.012 \pm 0.002$ & 299 & $0.03 \pm 0.00$ & $0.02 \pm 0.00$ & 9 \\
abaxlpi & 0.064 & 58298.5 & $18.394 \pm 0.005$ & $0.12 \pm 0.05$ & $0.047 \pm 0.004$ & 24 & $0.06 \pm 0.07$ & $0.05 \pm 0.06$ & 9 \\
abfgygp & 0.064 & 58322.0 & $18.032 \pm 0.003$ & $-0.01 \pm 0.02$ & $-0.076 \pm 0.003$ & 2 & $0.09 \pm 0.02$ & $0.07 \pm 0.01$ & 9 \\
abfhryc & 0.032 & 58323.9 & $16.468 \pm 0.003$ & $0.50 \pm 0.02$ & $-0.014 \pm 0.002$ & 99 & $0.06 \pm 0.05$ & $0.05 \pm 0.04$ & 11 \\
abimsyv & 0.088 & 58334.5 & $18.540 \pm 0.006$ & $1.04 \pm 0.04$ & $-0.046 \pm 0.004$ & $-5$ & $0.04 \pm 0.03$ & $0.06 \pm 0.03$ & 9 \\
abssuxz & 0.065 & 58377.0 & $18.155 \pm 0.008$ & $-2.10 \pm 0.10$ & $0.027 \pm 0.007$ & $-11$ & $0.05 \pm 0.07$ & $0.07 \pm 0.05$ & 4 \\

 \hline
\end{tabular}
\caption{SNe\,Ia with early excesses and their SALT3 properties from the \citet{Yao19} sample. Bump properties are quoted at a reasonable, BIC-maximizing $N$. $f_b$ refers to the maximum excess bump flux, whereas $f_{10}$ refers to the flux at 10 days post-explosion, in each band. Typical errors on $z$ and $t_0$ are less than $10^{-3}$ and $10^{-1}$, respectively.}
\label{table:yaoexcess}
\end{table*}

\begin{table*}[t]
\centering
\begin{tabular}{ |l|r|r|r|r|r|r|r|r|r| } 
  \hline
 
 ZTF ID & $z$ & $t_{exp}$ & $m_B$ (mag) & $x_1$ & $c$ & $\Delta$ BIC & $f_{b, r}/f_{10, r}$ & $f_{b, g}/f_{10, g}$ &  $N$\\ 
 \hline
 \multicolumn{10}{|c|}{Gold Tier} \\
 \hline
aayjvve & 0.047 & 58292.2 & $17.704 \pm 0.005$ & $-0.08 \pm 0.04$ & $0.065 \pm 0.004$ & 16 & $0.08 \pm 0.03$ & $0.06 \pm 0.02$ & 7 \\
abauprj & 0.024 & 58302.2 & $15.556 \pm 0.001$ & $1.34 \pm 0.01$ & $-0.033 \pm 0.001$ & 93 & $0.02 \pm 0.01$ & $0.02 \pm 0.01$ & 12 \\
 \hline
 \multicolumn{10}{|c|}{Bronze Tier} \\
 \hline
abkhcwl & 0.032 & 58344.3 & $16.426 \pm 0.003$ & $0.01 \pm 0.02$ & $-0.092 \pm 0.002$ & 8 & $0.00 \pm 0.10$ & $0.10 \pm 0.10$ & 8 \\
abuhzfc & 0.037 & 58383.0 & $17.239 \pm 0.004$ & $-0.78 \pm 0.06$ & $0.139 \pm 0.004$ & 2 & $0.02 \pm 0.02$ & $0.20 \pm 0.20$ & 8 \\
abvejbm & 0.044 & 58385.8 & $17.250 \pm 0.010$ & $-2.14 \pm 0.05$ & $-0.010 \pm 0.009$ & 3 & $0.05 \pm 0.04$ & $0.04 \pm 0.04$ & 8 \\
 \hline

 \hline
\end{tabular}
\caption{SNe\,Ia with early excesses and their SALT3 properties from the ZTF 2018 cosmology sample. Bump properties are quoted at reasonable, BIC-maximizing $N$. $f_b$ refers to the maximum excess bump flux, whereas $f_{10}$ refers to the flux at 10 days post-explosion, in each band.  SNe\,Ia ZTF18aayjvve and ZTF18abauprj are also found in the \citet{Yao19} sample.}
\label{table:dhawanexcess}
\end{table*}

\section{Results} \label{results}
\subsection{SNe\,Ia with Excesses}
 Unsurprisingly, due to the high cadence and exceptional pre-peak coverage, we find that most SNe\,Ia with significant, detectable early excesses come from the \cite{Yao19} sample.  In total, from the \cite{Yao19} sample, \cite{Burke22a} report 11 SNe\,Ia with early excesses: 2 gold, 4 silver, and 5 bronze. Our models (Section \ref{fitting}) find 14 SNe\,Ia (4 gold, 10 bronze\footnote{We note that these tiers are loosely modeled after \citet{Burke22a}, but use the criteria discussed in Section \ref{sec:tiers} and the Appendix.}) with excesses in the same sample, 7 of which overlap with the \citet{Burke22a} findings. One SN\,Ia (ZTF18aayjvve) is previously reported in \cite{Deckers22}; the other 6 SNe\,Ia (ZTF18abucvbf, ZTF18aasdted, ZTF18abaxlpi, ZTF18abfgygp, ZTF18aaslhxt, ZTF18abauprj), one of which we identify as a gold-tier excess, have not previously been reported. We find just 4 objects that can be classified as gold-tier no-excess SNe\,Ia due to the difficulty of ruling out bumps smaller than our smallest clear detection in the three-day cadence ZTF data.

With the same systematic criteria, we search for early excesses in the SNe\,Ia light curves from the ZTF 2018 cosmology data release. From the \cite{Dhawan22} sample, we identify 2 gold-tier and 3 bronze-tier excesses, 2 of which overlap with the excesses found in the previously discussed sample; ZTF18abvejbm and ZTF18abuhzfc are unique to the \citet{Dhawan22} sample. We find only two objects that can be classified as gold-tier no-excess SNe\,Ia.  

For the other early-excess SNe\,Ia found in \cite{Burke22b} or \cite{Yao19} with light curves also available in the \citet{Dhawan22} dataset, we find that the \citet{Dhawan22} data allowed detection of early excesses in cases where the original ZTF forced photometry light curves were often too poorly sampled to detect them. For SNe that we do not detect excesses in, but have been reported as having early excesses in other analyses, we remove them our list of gold-tier non-detections. We find $\sim$1 gold-tier early excess and $\sim 5-10$ bronze-tier early excesses in the ZTF 2019-2022 sample, but ultimately do not include these SNe\,Ia due to the small number of excess detections and some data quality issues that affected how confident we could be in those detections, including lack of pre-explosion data, template over-subtraction, significant light-curve outliers, and under-estimated uncertainties (it is likely that these light curves will be improved in future data releases).

 Overall, SNe\,Ia with early excesses have relatively similar lightcurve shapes $x_1$ and colors $c$ compared to SNe\,Ia without them. From the \citet{Yao19} sample, early-excess SN\,Ia have $\overline{x_1} = -0.100 \pm 0.005$  and $\overline{c} = 0.0180 \pm 0.0007$, compared to excess-free SN, $\overline{x_1} = -0.798 \pm 0.04$ and $\overline{c} = 0.0308 \pm 0.0005$. 

\subsection{Hubble Residuals}
\label{sec:hubbleres}

\begin{figure}
     \centering
     \subfloat{
         \centering
         \includegraphics[width=0.46\textwidth]{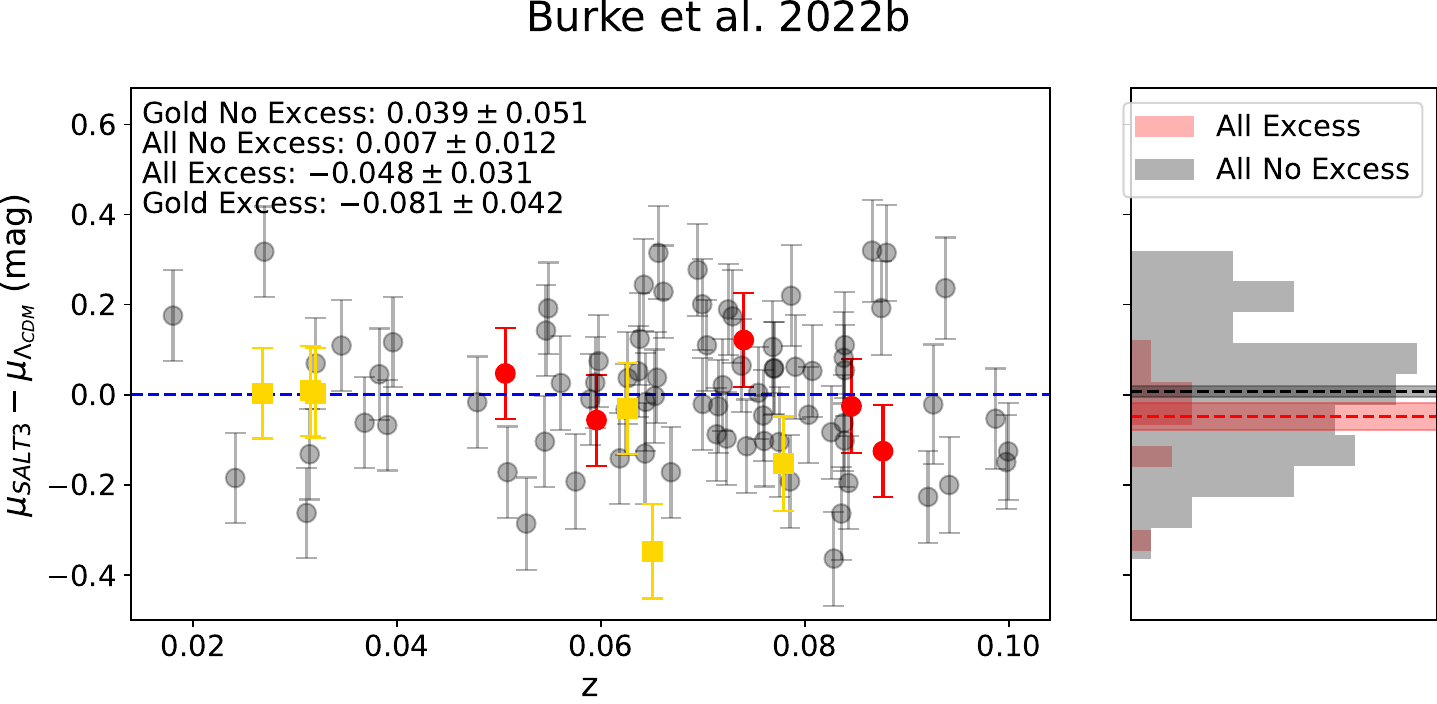}}
         \hfill
    \subfloat{
         \centering
         \includegraphics[width=0.46\textwidth]{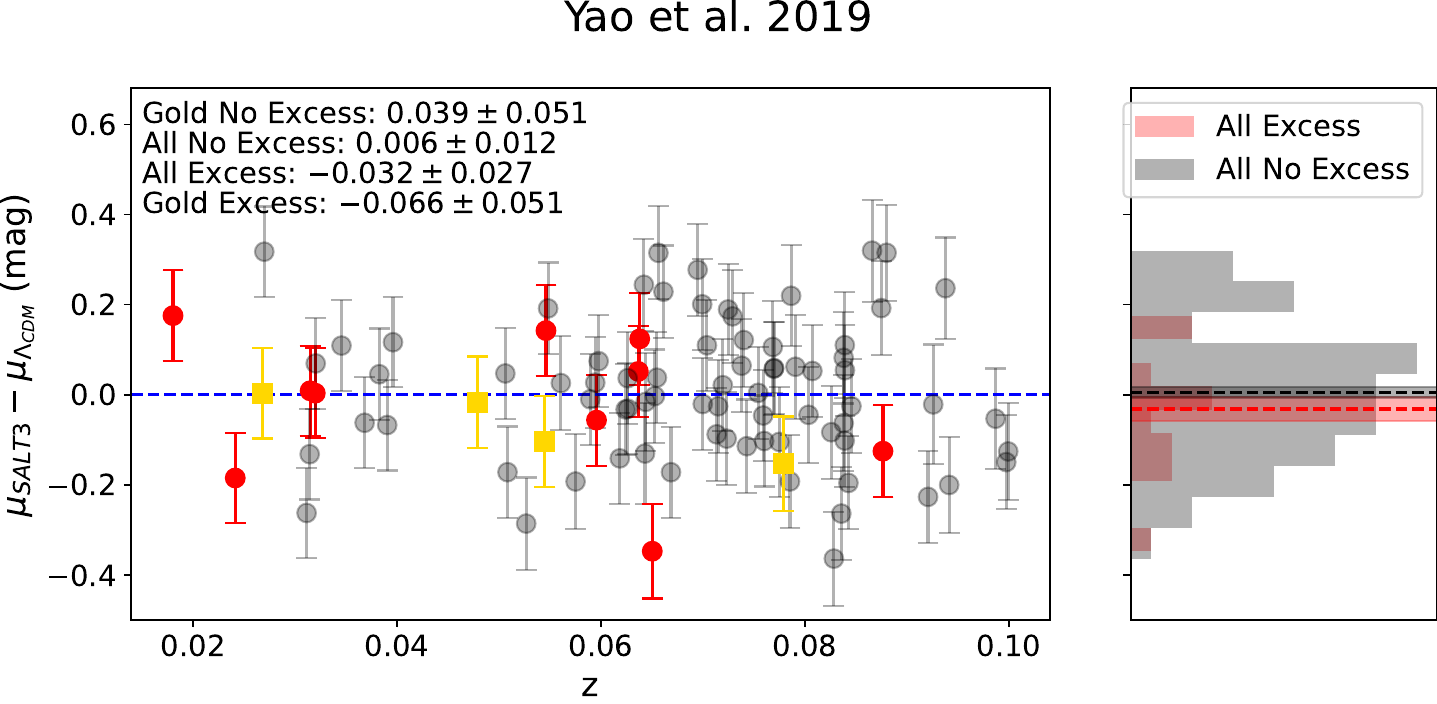}}
         \hfill
     \subfloat{
         \centering
         \includegraphics[width=0.46\textwidth]{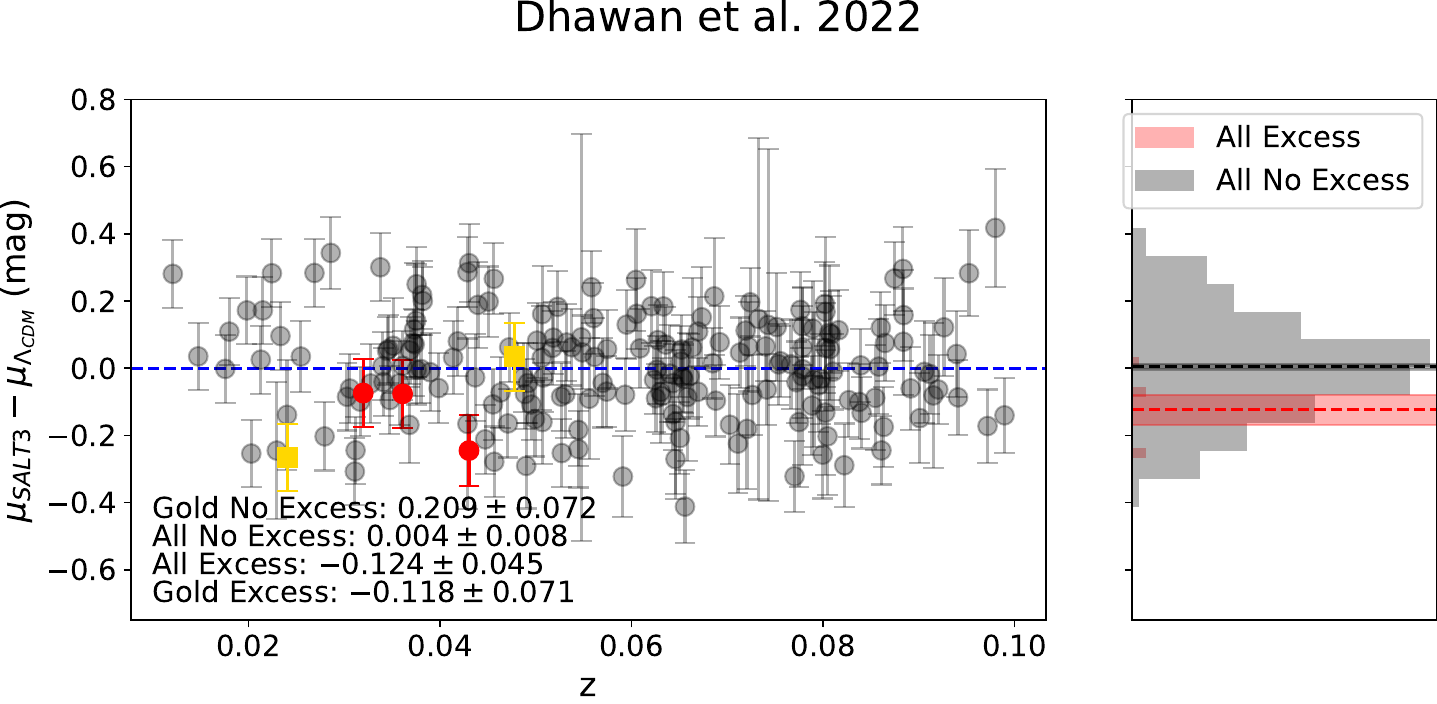}}
     \hfill
        \caption{Hubble residuals for our sample. (Top): SNe\,Ia from \citet{Burke22b}, (Middle): SNe\,Ia from \citet{Yao19}, and (Bottom): SNe\,Ia from \citet{Dhawan22}. For all panels, SNe\,Ia in grey are non-excess SNe\,Ia, red are bronze excess, and gold are gold excess. The histograms to the right of each panel show the Hubble residual dispersion for all excess-having (gold $+$ bronze) SNe\,Ia in red and excess-less (any other object) SNe\,Ia in gray;  average and 1$\sigma$ intervals are shown with the shaded bars.}

        \label{fig:hubble}
    
\end{figure}

Flux excesses may originate from different SN\,Ia progenitor systems and/or explosion mechanisms such that the cosmological standardization of the SNe\,Ia with excesses may be adversely affected.
Therefore, it is reasonable to expect that there may be a resulting difference in the measured distance moduli, $\mu$, between the excess-having and no-excess SN\,Ia populations. After fitting for Tripp equation nuisance parameters (Section \ref{sample}), including corrections for 
light-curve shape, color and host-galaxy mass, and marginalizing over redshift-dependent trends in a manner similar to \citet{Marriner11}, we compare Hubble residuals ($\mu_{SALT3} - \mu_{\Lambda_{CDM}}$) for each sample; results are shown in Figure \ref{fig:hubble}.

We begin with the literature sample of SNe\,Ia with early excesses identified in \cite{Burke22b}. We combine the \citet{Burke22b} ``gold" and ``silver" classifications into a single ``gold" tier, and add in the \citet{Burke22b} ``bronze" class for the full population of excess-having SN. The gold-tier no-excess SNe\,Ia are taken from our analysis of \cite{Yao19} light curves, as discussed in Section \ref{sec:tiers}. The combined gold-tier and bronze-tier samples, representing all excess-having SNe, when compared to all other SNe, have
$\Delta {\rm HR} = -0.055 \pm 0.033$~mag, or about $1.7\sigma$. Within the gold tier, excess-having SNe\,Ia and the gold-tier no-excess sample have comparable Hubble residuals
$\Delta {\rm HR} = -0.120 \pm 0.066$~mag, although we caution that comparisons within the gold tier are difficult due to low statistics (4 excess and 4 no-excess SNe). 

Next, using the same data source as \cite{Burke22a}, detailed in \cite{Yao19}, we use the results of our systematic search for early excesses as described in \ref{sec:tiers}.  We define new gold and bronze tiers, and find that our identification of SNe\,Ia with early excesses is somewhat more conservative than \cite{Burke22a} (see Section \ref{sec:discussion} for a discussion of the different samples in this work).  We again find that, on a population level, excess-having SNe\,Ia have lower Hubble residuals (and are thus brighter); we measure $\Delta {\rm HR} = -0.038 \pm 0.030$~mag. This sample shares the same gold-tier non-detection sample (4 SNe) as the \cite{Burke22a} paper; within this more limited sample, we measure $\Delta {\rm HR} = -0.105 \pm 0.072$.

Finally, with SNe\,Ia from the ZTF 2018 cosmology sample, we again find a similar trend. All supernovae with excesses are on average marginally brighter than those without excesses, with $\Delta {\rm HR} = -0.128 \pm 0.046$~mag. This trend also emerges when just looking at the gold tier:  $\Delta {\rm HR} = -0.327 \pm 0.102$~mag. 

Taking the no-overlap weighted average of the \citet{Yao19} and \citet{Dhawan22} SNe\,Ia samples, we find that gold-tier excess-having and excess-less SNe\,Ia differ in Hubble residuals by $\Delta {\rm HR} = -0.198 \pm 0.062 = 3.2 \sigma$, and all (gold and bronze) excess-having SNe\,Ia differ from the remaining SNe\,Ia by $\Delta {\rm HR} = -0.056 \pm 0.026$~mag, or $2.2 \sigma$. This suggests that SNe\,Ia with detected early excesses may be slightly intrinsically brighter at maximum, after shape, color, and host-galaxy mass correction, by $\sim$0.05--0.1~mag, compared to those without excesses.  We note that the $\Delta {\rm HR}$ sizes are similar when the mass step is not included, with \citet{Yao19} slightly smaller before mass step correction and \citet{Dhawan22} slightly larger.

\subsection{Host-Galaxy Masses}
\begin{figure}
     \centering
     \subfloat{
         \centering
         \includegraphics[width=0.46\textwidth]{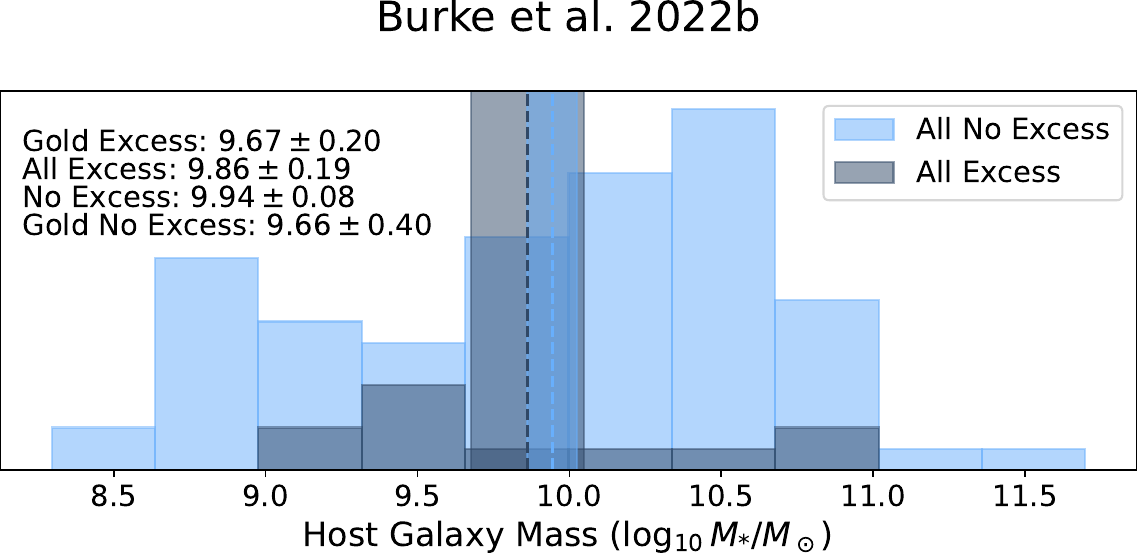}}
    \hfill
    \subfloat{
         \centering
         \includegraphics[width=0.46\textwidth]{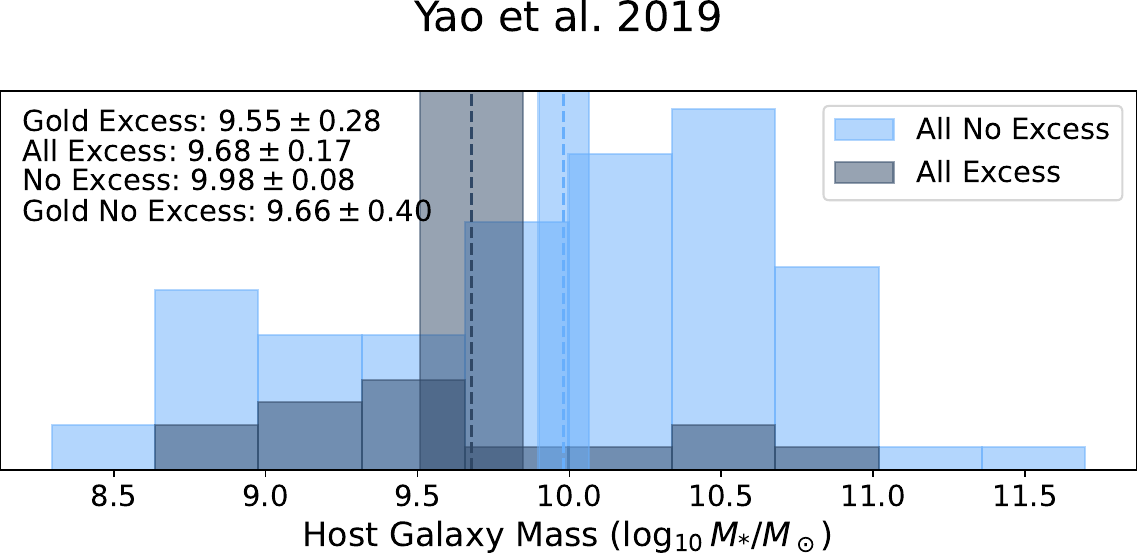}}

    \subfloat{
         \centering
         \includegraphics[width=0.46\textwidth]{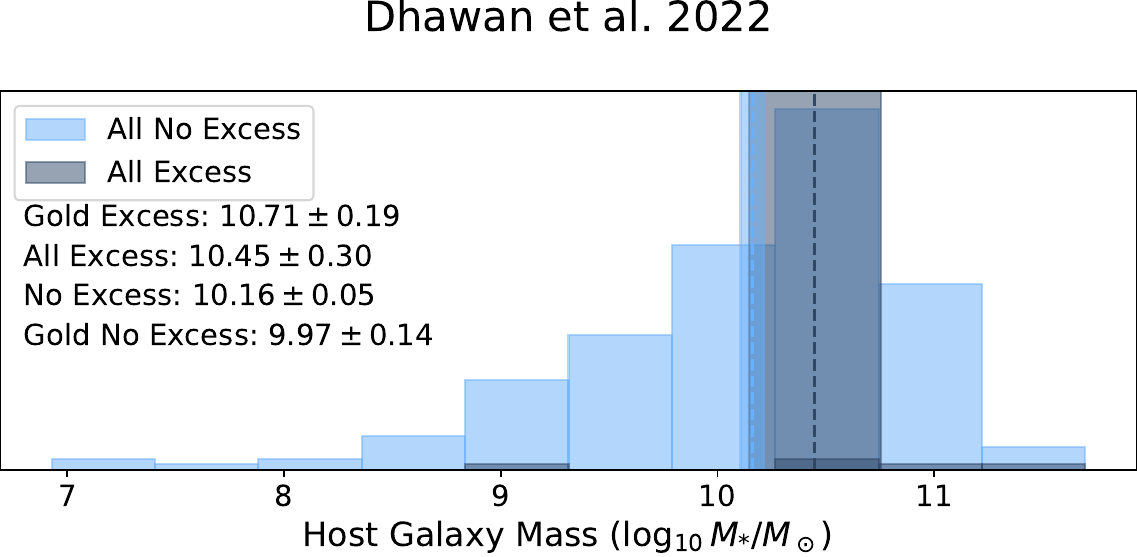}}
     \hfill
        \caption{Host-galaxy masses for all excess-having (gold + bronze) and all no-excess (all other objects) SNe\,Ia, unscaled. Mean and $1\sigma$ intervals are indicated in the same color scheme by dashed line and shaded vertical bars, respectively.  The mass distributions of the \citet{Yao19} data (top and middle panels) versus the \citet{Dhawan22} sample (bottom panel) appear significantly different, perhaps due to the host-galaxy redshift cut in \citet{Dhawan22}.
        }
        \label{fig:masses}
    
\end{figure}

Many possible physical explanations for the source of early time excesses in SNe\,Ia --- for example, single-degenerate progenitor systems \citep{Kasen10} and $^{56}$Ni distribution \citep{Piro13} --- would be expected to cause an increased early excess rate in certain host-galaxy environments given the predicted change in delay times between different SN\,Ia progenitor models (see \citealp{Maoz14} for a review).  
For that reason, we compare the stellar masses ($\log_{10} M_* / M_\odot$) of matched host galaxies (see Section \ref{sample}) for each sample. Results are shown in Figure \ref{fig:masses}.

Starting with the SNe\,Ia with early excesses identified in \cite{Burke22a}, we find that excess-having supernovae, represented by the combined (gold and bronze) sample, do not exhibit any difference in host-galaxy masses as compared to excess-less (all other) SNe; $\overline{M_{e}} = 9.86 \pm 0.19$~dex and $\overline{M_{n}} = 9.94 \pm 0.08$~dex, with $\overline{M_{e}} -\overline{M_n} = -0.08 \pm 0.21$~dex. Within the gold tier, though, we again find that SNe\,Ia with excesses are found in galaxies with comparable mass, with the average ($N = 5$) $\log_{10} M_*/M_\odot \equiv \overline{M}_{e} = 9.67 \pm 0.20$~dex, compared to the average ($N = 4$) $\overline{M}_{n} = 9.66 \pm 0.40$~dex for gold-tier excess-less SN; $\overline{M_e} - \overline{M_n} = 0.01 \pm 0.45$~dex. 

Based on the SNe\,Ia in \cite{Yao19}, we find that the combined sample of excess-having SNe\,Ia actually has a subtle preference for lower-mass galaxies, with $\overline{M_{e}} = 9.68 \pm 0.17$~dex, $\overline{M_{n}} = 9.98 \pm 0.08$~dex, and $\overline{M_{e}} -\overline{M_n} = -0.30 \pm 0.19$~dex (1.6$\sigma$). Gold-tier SNe\,Ia, appear to exhibit no preference for lower-mass galaxies; $\overline{M_{e}} = 9.55 \pm 0.28$~dex, $\overline{M_{n}} = 9.66 \pm 0.40$~dex, for a difference of $\overline{M_{e}} -\overline{M_n} =  -0.11 \pm 0.49$~dex. 

Finally, with SNe\,Ia from the ZTF 2018 cosmology data release, we find that, on average, the excess-having supernovae appear in slightly higher-mass galaxies than those without excesses: $\overline{M_e} = 10.45 \pm 0.30$~dex, $\overline{M_n} = 10.16 \pm 0.05$~dex, and $\overline{M_e} - \overline{M_n} = 0.29 \pm 0.30$~dex. This trend holds for the gold tier: $\overline{M_n} = 9.97 \pm 0.14$~dex and the gold-tier excess-having sample has $\overline{M{e}} = 10.71 \pm 0.19$, with $\overline{M_{e}} -\overline{M_n} = 0.74 \pm 0.24$~dex. 

Overall, there is no clear trend between early-excess supernovae and the masses of their host galaxies. When taking the no-overlap weighted average of the \citet{Yao19} and \citet{Dhawan22} samples, we find that there is an insignificant average offset of $-0.25 \pm 0.17$~dex in the host galaxy mass of SNe\,Ia with and without flux excesses. However, we note that because \citet{Dhawan22} only uses samples with measured host-galaxy redshifts, there is a significant bias in the host-galaxy masses of this sample.  While we wouldn't necessarily expect this selection to change the relative difference in host-galaxy mass between the excess-having and no-excess populations, the \citet{Yao19} sample is less biased as a function of host mass and has better sample statistics.


\subsection{Light-curve Properties}

\begin{figure*}
     \centering
     \subfloat[Power-law slopes, $\alpha$, in ZTF $g$ and $r$ bands.  Green and red bars span the 1$\sigma$ uncertainty on the mean for $g$ and $r$ band, respectively.]
     {
         \centering
         \includegraphics[width=0.45\textwidth]{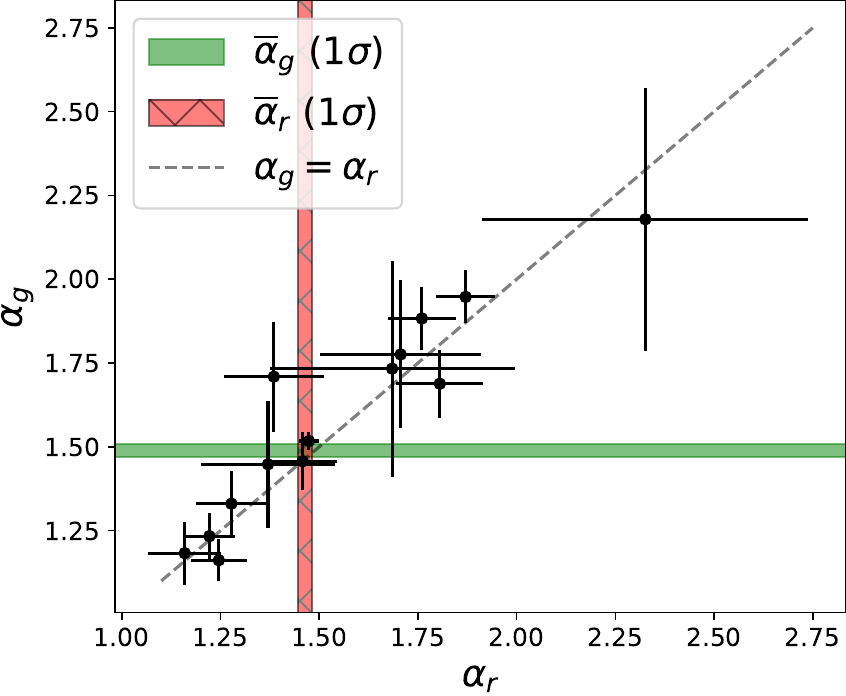} 
    }
         \hfill
    \subfloat[Traces of relative excess sizes, scaled by 10-day flux, in $g$ and $r$ bands.  The gray bar spans the 1$\sigma$ uncertainty on the mean.]{
         \centering
         \includegraphics[width=0.45\textwidth]{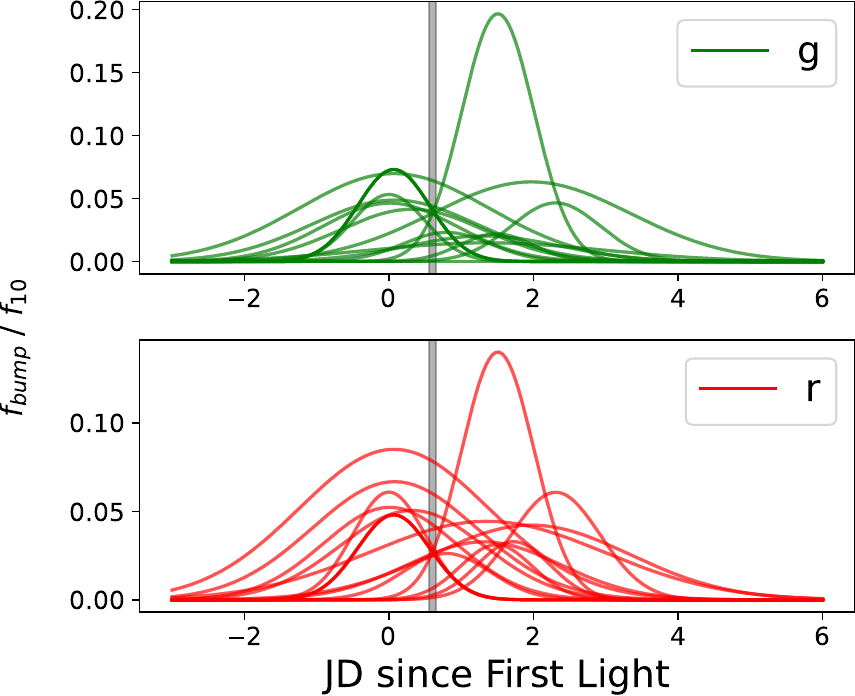}}
         \hfill
    \subfloat[Relative excess sizes in $g$ and $r$ bands.  Green and red bars span the 1$\sigma$ uncertainty on the mean for $g$ and $r$ band, respectively.]{
         \centering
         \includegraphics[width=0.45\textwidth]{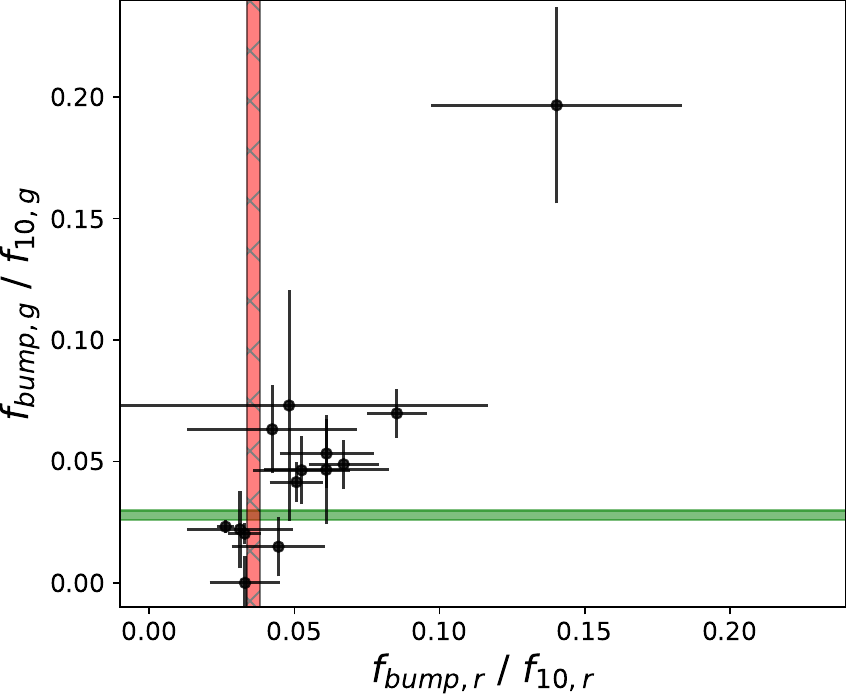}}
     \hfill
        \caption{Properties of light curves from \cite{Yao19} with early excesses. $1\sigma$ uncertainty regions are shaded.}
        \label{fig:lightcurveprops}
    
\end{figure*}
Finally, we analyze the statistical properties of the light curves of early-excess SNe\,Ia from the \cite{Yao19} sample (see Figure \ref{fig:lightcurveprops}). All averages quoted in this section are weighted by the uncertainties. Based on a fireball model \citep{Riess99}, we expect the first few days post-explosion to be well-fit by a power law, approximately $t^2$, but previous studies have found a diversity of power-law slopes \citep{Miller20,Fausnaugh21}. Among 14 total SNe\,Ia with early excesses, we find $\alpha_r = 1.46 \pm 0.02$ and $\alpha_g = 1.49 \pm 0.02$, consistent with individual SNe\,Ia from \citet{Miller20} but inconsistent with their population average ($2.01 \pm 0.02$ in the $r$ band), indicating that the rise rates may be less steep for early excess SNe. A different slope as a function of filter would be expected for a changing SED temperature, but here we find that 
$g$-band rises are consistent with $r$-band rises with a difference of $0.03 \pm 0.03$.


While many physical models of early excesses predict stronger excesses at bluer wavelengths, we do not find a significant color dependence.
Excess-having SNe\,Ia from the literature also have a range of colors, though many of the earliest known examples were blue: 2017cbv had a $g-r$ color of approximately $-0.2$~mag \citep{Hosseinzadeh17}, as did 2018oh \citep[albeit very uncertain]{Dimitriadis19}.  SN~2012cg had $B-V \simeq -0.1$~mag in its earliest epoch.  Here, we see a wider range of colors; for example, we find an exclusively red excess in ZTF18aaxsioa and an average $g-r = 0.06 \pm 0.09$~mag for the other 13 SNe\,Ia.  Other studies of ZTF SNe have found similar results, with 
\citet{Deckers22} have $g-r$ colors spanning from $-0.35$ to $+0.15$ mag, though with an average of $-0.14$~mag (uncertainties range from $\sim$0.1--0.6~mag).  Similarly, \citet{Burke22b} saw that ZTF SNe with early excesses often have redder early time colors (\citealp{Burke22a} also saw this in a literature sample using $B-V$ colors).

Averaged among all \citet{Yao19} SNe\,Ia with detections, the early excess flux has mean absolute magnitudes of $M_g = -15.55 \pm 0.33$ and $M_r = -15.72 \pm 0.25$.  Within our (rather restrictive) priors, we find that the average early excess peaks at $0.60 \pm 0.05$ days after explosion, and has a Gaussian standard deviation of $\sigma = 0.71 \pm 0.02$ days.


\section{Discussion}
\label{sec:discussion}

\subsection{Robustly Identifying Early Excess SNe\,Ia}
\label{sec:excess_discussion}

The task of systematically selecting early excess SNe\,Ia is a difficult one, due to non-uniformity in both the input light curves and the diverse nature of the early excesses. We found in this work that the detection of early excesses can be degenerate with the slope of the power-law rise and the time of explosion (Figure \ref{fig:multinest}).  Lack of sufficient sampling immediately post-explosion, outliers in the light curves, and occasional under-estimated uncertainties added additional complexity.  Because early excess detections are often subtle, we found that detecting them is not always unambiguous; we did not confidently find excesses for several objects with early excesses reported in the literature.  In this study, following \citet{Burke22b}, we felt this necessitated including different tiers of confidence for how reliable the detections were and several criteria to account for uncertainties in the fitting and data.  While future data will inevitably increase the statistics for early excess studies, we expect that high-cadence survey or follow-up data will be particularly valuable for robust detections.

\subsection{Subsample Comparison and Future Improvements}

While the \citet{Dhawan22} data were intended to be used for cosmological analyses, the ZTF forced-photometry data from \citet{Yao19} were not.  Despite this, the light-curve calibration appears to be sufficient to yield a Hubble-diagram scatter that is competitive with other recent cosmology samples, including \citet{Dhawan22}.

We note that we found two differences between the \citet{Yao19} and \citet{Dhawan22} samples in our distance fitting. The first is that the $\beta$ parameter relating color to distance modulus is different at just under 3$\sigma$ significance between the two samples (Table \ref{table:tripp}).
This difference could be related to either sample selection effects or uncorrected calibration uncertainties in the 2018 ZTF photometry.

We note that both subsamples were derived with different photometric pipelines. While the \citet{Yao19} used the difference images from IPAC \citep[e.g., see][]{Masci19} and PSF photometry with \texttt{ForcedPhotZTF}, \citet{Dhawan22} construct custom template images and difference images and then used \texttt{photutils} \citep{larry_bradley_2023_7946442} for performing forced photometry. Given these differences in the processing,  we do not combine the photometric data, instead analyzing each self-consistent sample independently and using weighted averages to yield final constraints.

Secondly, the distance bias significantly differs between \citet{Yao19}-derived distances and those of \citet{Dhawan22}.  This is likely due to sample selection; while \citet{Dhawan22} started with a magnitude-limited sample and made additional cuts on host-galaxy redshift availability (thereby favoring bright host galaxies), \citet{Yao19} selected their sample based on the availability of high-S/N, high-cadence early-time light curves.  We attempt to marginalize over these differences by fitting a model to the redshift-dependent Hubble residuals and correcting for the trends, which are substantially different between the two data sets in this work.  This procedure is standard practice for SN\,Ia cosmology \citep[e.g.,][]{Marriner11}, but cosmology analyses also typically use large Monte Carlo simulations to estimate and correct for distance bias, which we have not performed here because it is not necessary for us to separate redshift-dependent Malmquist biases from differences in the cosmological model.  More complex evaluations of the distance bias will likely be performed by the ZTF team in the future, which could include pixel-level pipelines to model search/classification efficiencies and perform scene-modeling photometry.

Despite those caveats, our methodology avoided most major sources of systematic uncertainty by performing self-consistent comparisons between excess-having and no-excess SNe\,Ia within each individual data set.  Higher-dimensional models of selection effects (i.e., correcting for $x_1$ and $c$ measurement biases; \citealp{Kessler17}) would be unlikely to substantially change the results given the high-S/N ZTF data and the similar average values of $x_1$ and $c$ between the samples.  We consider it more likely than not that SN\,Ia distance measurements do correlate with behavior in the early time light curves of SNe\,Ia, and we anticipate future work on this topic with larger, better-calibrated, cosmologically useful SNe\,Ia samples such as future ZTF data releases.

\subsection{Possible Impact on Measurements of Dark Energy Properties}

\begin{table}[ht]
\centering
\begin{tabular}{ |c|c|c|c|} 
 \hline
  $\Delta {\rm HR}$ (mag)&$\Delta f_{e}$ (\%)&$\mu$ bias (mag)&$\Delta w\ (z = 1)$\\
 \hline
0.05&5&0.003&0.007\\
0.05&10&0.005&0.013\\
0.05&15&0.007&0.020\\
0.10&5&0.005&0.013\\
0.10&10&0.010&0.027\\
0.10&15&0.015&0.041\\
0.15&5&0.007&0.020\\
0.15&10&0.015&0.041\\
0.15&15&0.022&0.062\\
 \hline
\end{tabular}
\caption{Biases on $w$ inferred from comparing a $z = 1$ SN\,Ia sample to a $z = 0.05$ sample as a function of different Hubble residual steps due to early excesses ($\Delta {\rm HR}$) and changes in early excess fraction between $z = 0$ and $z = 1$ ($\Delta f_{e}$).  The product of the first and second columns is the average bias on distance at $z = 1$ ($\mu$ bias).}
\label{table:w}
\end{table}

Although early-excess SNe\,Ia  are a minority of the SN\,Ia sample, it is possible that they comprise a larger fraction of high-redshift samples.  Here, we see hints that they might be more prevalent in low-mass galaxies, which are likely those with younger stellar populations.  Additionally, if they are indeed slightly brighter, surveys are more likely to detect them near the high-$z$ magnitude limit.

As a back-of-the-envelope example, in the Pantheon$+$ sample \citep{Scolnic22}, 58\% of the SNe\,Ia at $z < 0.1$ but just 25\% of SNe\,Ia at $z > 0.8$ have host-galaxy masses greater than 10~dex.  Early excesses from \citet{Yao19} are present in $\sim$9\% of the mass $>10$~dex SN\,Ia sample, but 23\% of the $<$10~dex sample; this equates to 15\% of the full low-$z$ sample and 20\% of the $z > 0.8$ sample.  If early excess SNe\,Ia have 10\% brighter Hubble residuals, this could imply a systematic offset in the distances of $\sim$0.5\% between high- and low-redshift SNe.
Such an offset could introduce 1-2\% biases on the dark energy equation of state parameter, $w$, which would be a dominant systematic in current $w$ measurements \citep{Brout22}.  Table \ref{table:w} demonstrates how changes in excess fraction with redshift and different values of $\Delta$HR (chosen to be consistent with our results) would naively translate to biases on $w$ for a $z = 1$ SN\,Ia sample.

This rough calculation highlights the importance of investigating these potential changes in the SN\,Ia population over cosmic time.  It is not intended as a definitive assessment of what biases might exist, and it assumes that the early excess rate as a function of host mass is fixed -- likely a poor assumption.  For example, the SN\,Ia rate as a function of host-galaxy mass \citep{Brown19,Wiseman21} appears to increase in low-metallicity environments due to an increased binary fraction \citep{Gandhi22,Johnson23}, which could affect e.g., single-degenerate progenitors to a different degree than double-degenerate progenitors.
It is also not necessary to observe early excesses at high redshift to correct for this bias; for example, a redshift-binned host-galaxy mass correction, would likely correct for this type of systematic distance offset (use of twin or sibling SNe --- SNe with similar spectra or in similar host-galaxy types --- could be another option; e.g., \citealp{Fakhouri15,Hoogendam23}).  However, current analyses do not make such corrections, and without {\it a priori} knowledge of what systematics could be problematic, these types of cosmological biases could be neglected.

\section{Conclusions}
In this work, we search for a change in Hubble residuals between SNe\,Ia with and without flux excesses in their early-time light curves.  Although there is no consensus physical explanation for why a minority of SNe\,Ia have rises that are not well represented by a single power-law model, we hypothesize that such behavior could be correlated with intrinsic SN\,Ia luminosity differences that propagate to distance measurements.

By examining two samples of ZTF SNe\,Ia with well-sampled early time light curves, we find 17 SNe\,Ia with significant detected excesses.  These SNe\,Ia have modest evidence for brighter Hubble residuals, $\Delta HR = -0.056 \pm 0.026~{\rm mag}$ (2.2$\sigma$ significance), compared to the rest of the sample.  SNe\,Ia with early excesses are found more often in lower-mass host galaxies, though the statistical significance is low, and host-galaxy selection effects are inherent in one of our samples.  Our detected excesses have an average $g-r$ color of $0.06 \pm 0.09$~mag.  We also find that excess-having SNe\,Ia rise less quickly than the mean \citet{Yao19} SN\,Ia as measured by \citet{Miller20}.

This work highlights the importance of connecting new diagnostics of SN\,Ia physics to cosmological parameter measurements.  Fortunately for the current study, future data releases from ZTF and ATLAS \citep{Tonry18}, alongside hundreds of thousands of SN\,Ia discoveries from the Rubin Observatory, will give a clear indication of the degree to which early time excesses correlate with distance measurements.  However, future surveys such as the {\it Roman Space Telescope} will observe SNe\,Ia at unprecedented redshifts ($z > 2$) to measure dark energy, and therefore require a comprehensive understanding of the ways that SN\,Ia distance measurements could become biased as their progenitors evolve with cosmic time.



\label{conclusions}

\software{
{\tt Astropy} \citep{Astropy13:paperI, Astropy18:paperII},
{\tt sncosmo} \citep{barbary21},
{\tt GHOST} \citep{Gagliano21}, {\tt dustmaps} \citep{Green2018},
{\tt SciPy} \citep{2020SciPy-NMeth},
{\tt matplotlib} \citep{Hunter07},
{\tt NumPy} \citep{Harris20},
{\tt PyMultiNest} \citep{Buchner14},
{\tt corner} \citep{corner}
}

\begin{acknowledgements}
C.Y. and D.O.J acknowledge support from {\it HST} grants HST-GO-17128.028 and HST-GO-16269.012 awarded by the Space Telescope Science Institute (STScI), which is operated by the Association of Universities for Research in Astronomy, Inc., for NASA, under contract NAS5-26555.
\end{acknowledgements}

\appendix 
\section{Early-Excess Criteria}\label{sec:appendix}

\begin{figure*}
  \centering 
     \subfloat{
         \centering
         \includegraphics[width=0.48\textwidth]{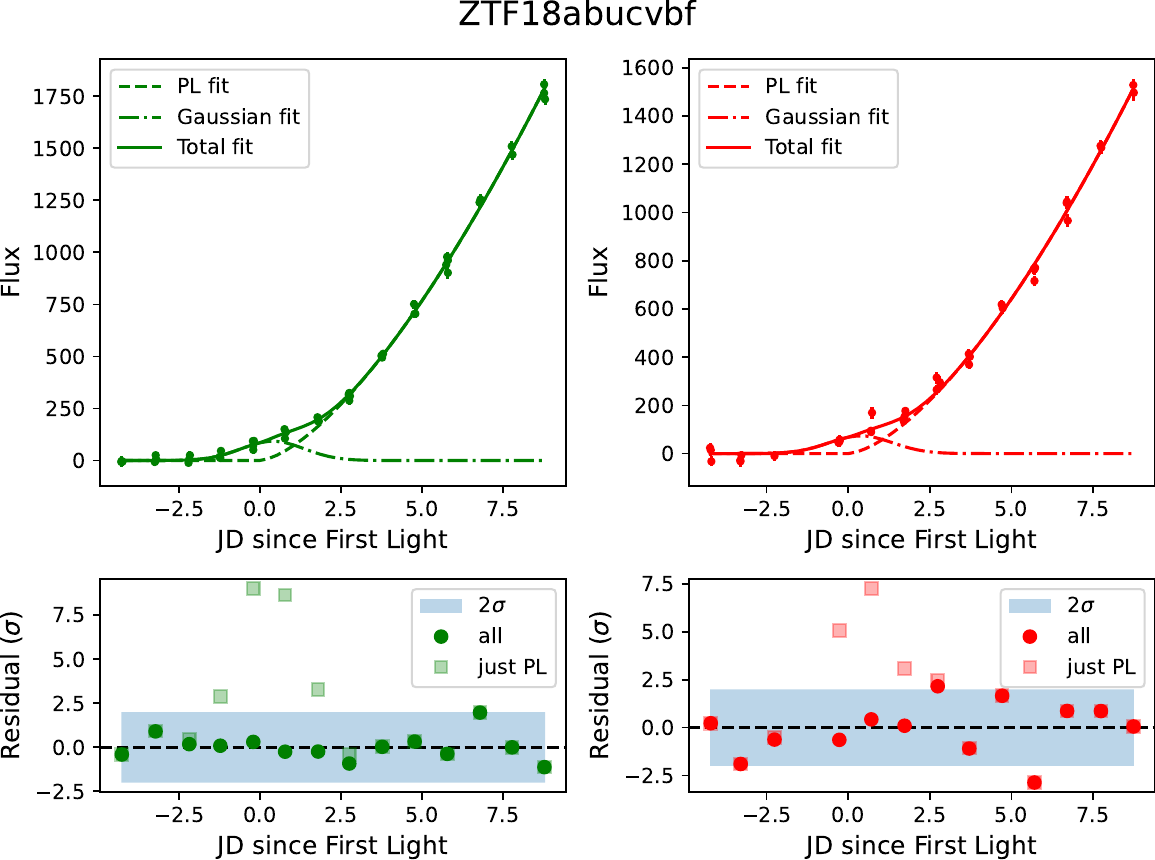}}
         \hfill
    \centering
     \subfloat{
         \centering
         \includegraphics[width=0.48\textwidth]{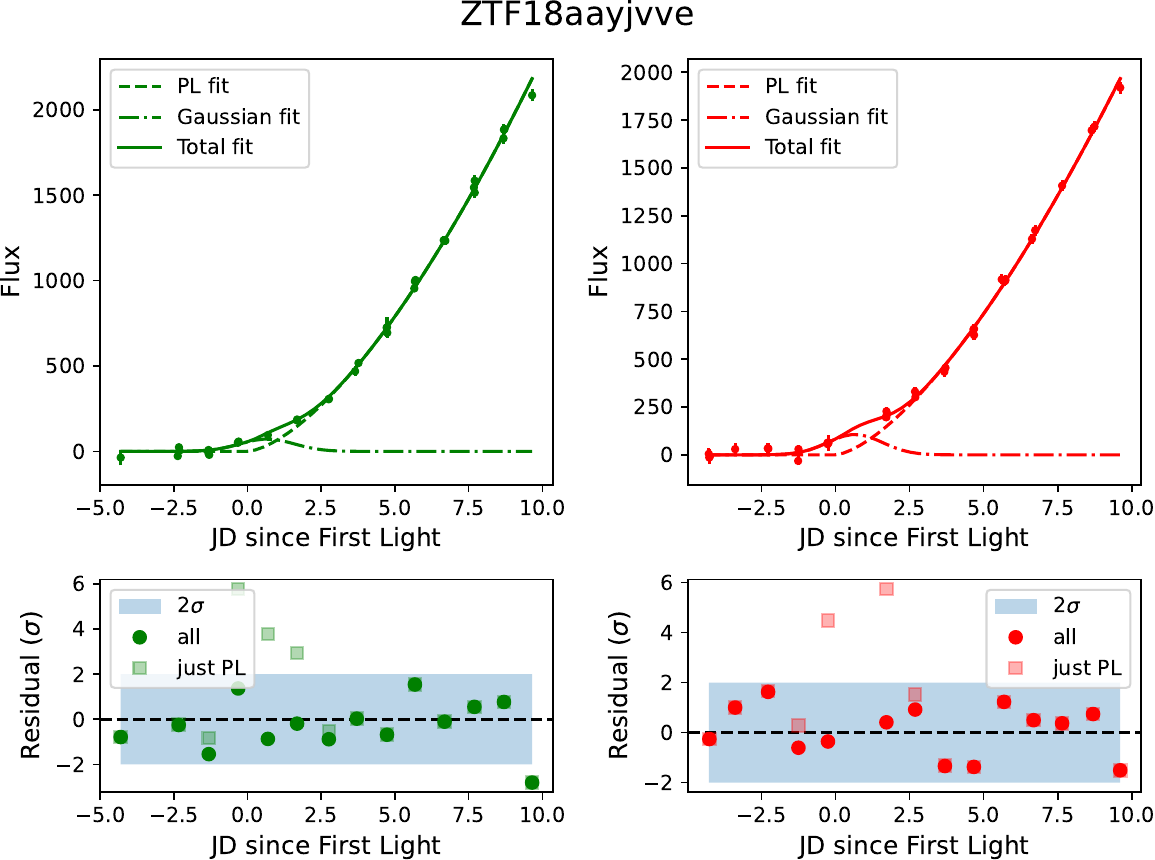}}
        \hfill
    \centering
     \subfloat{
         \centering
         \includegraphics[width=0.48\textwidth]{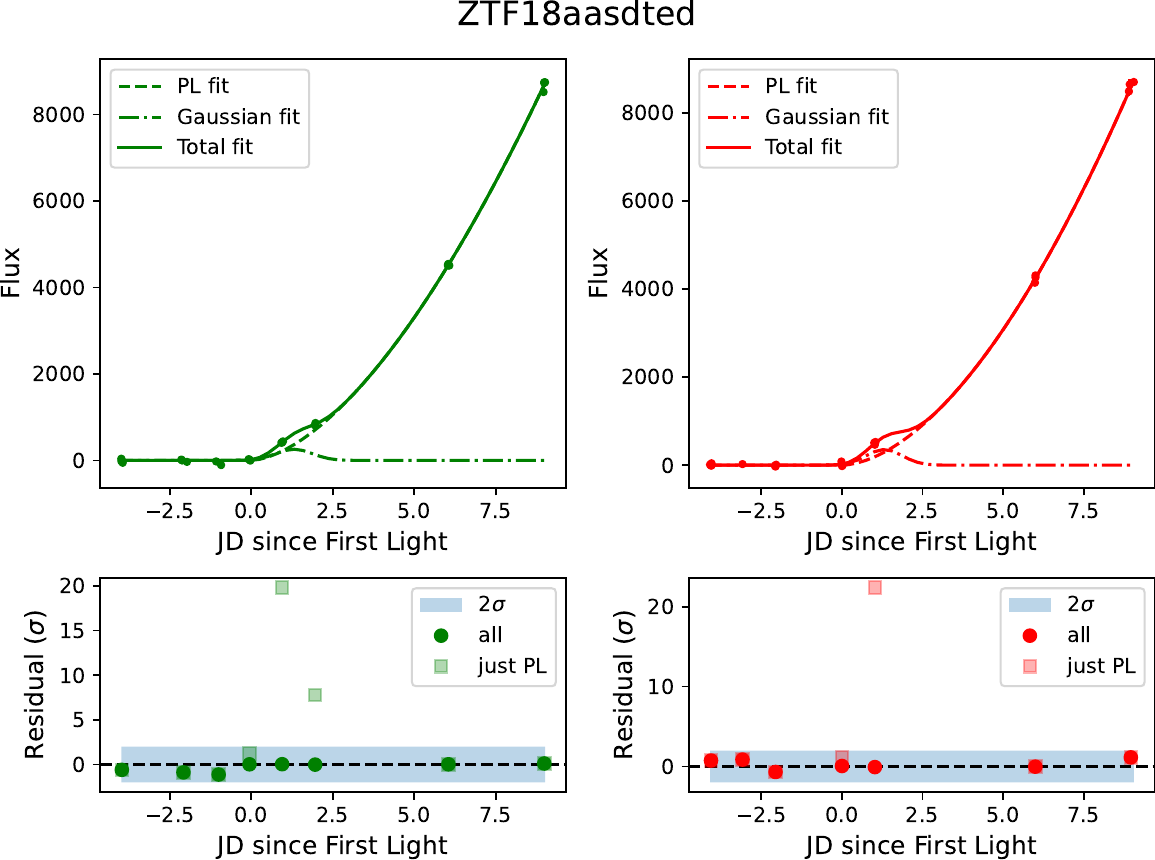}}
         \hfill
    \centering
     \subfloat{
         \centering
         \includegraphics[width=0.48\textwidth]{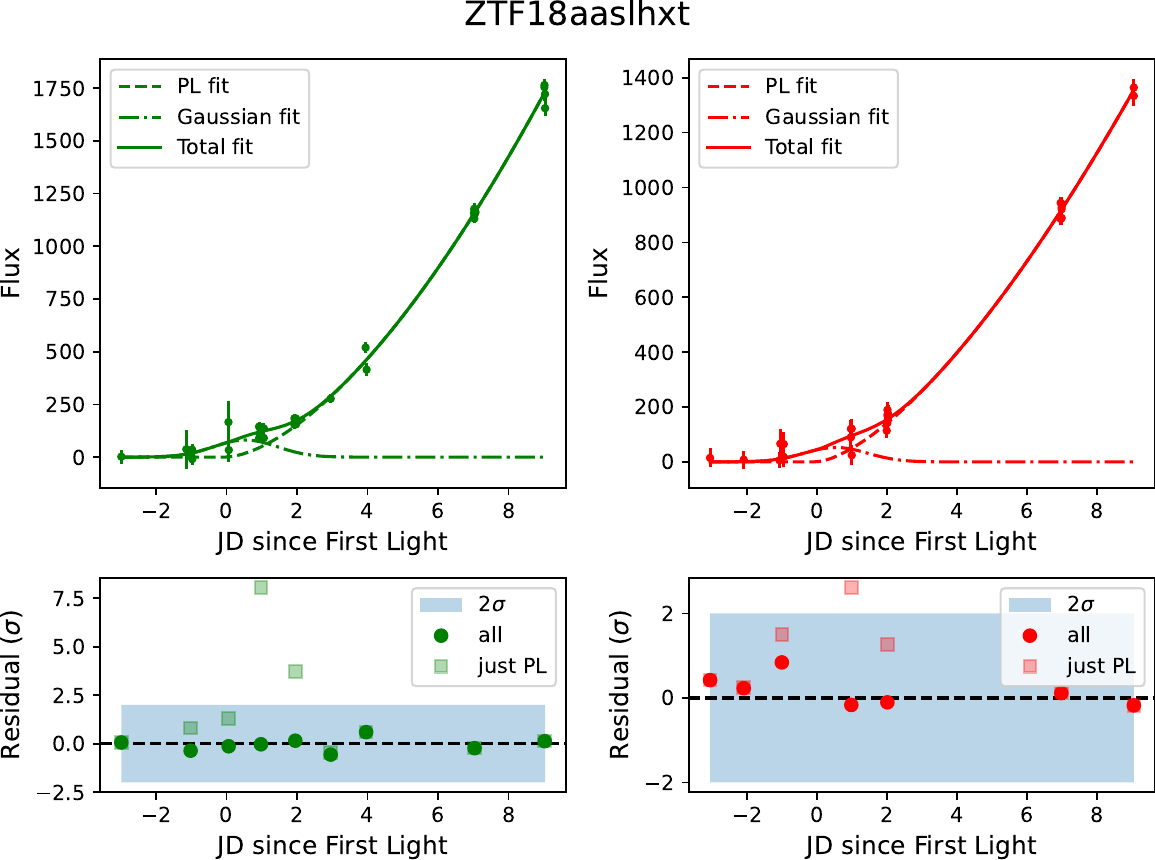}}
         \hfill
     \hfill
     \centering
     \subfloat{
         \centering
         \includegraphics[width=0.48\textwidth]{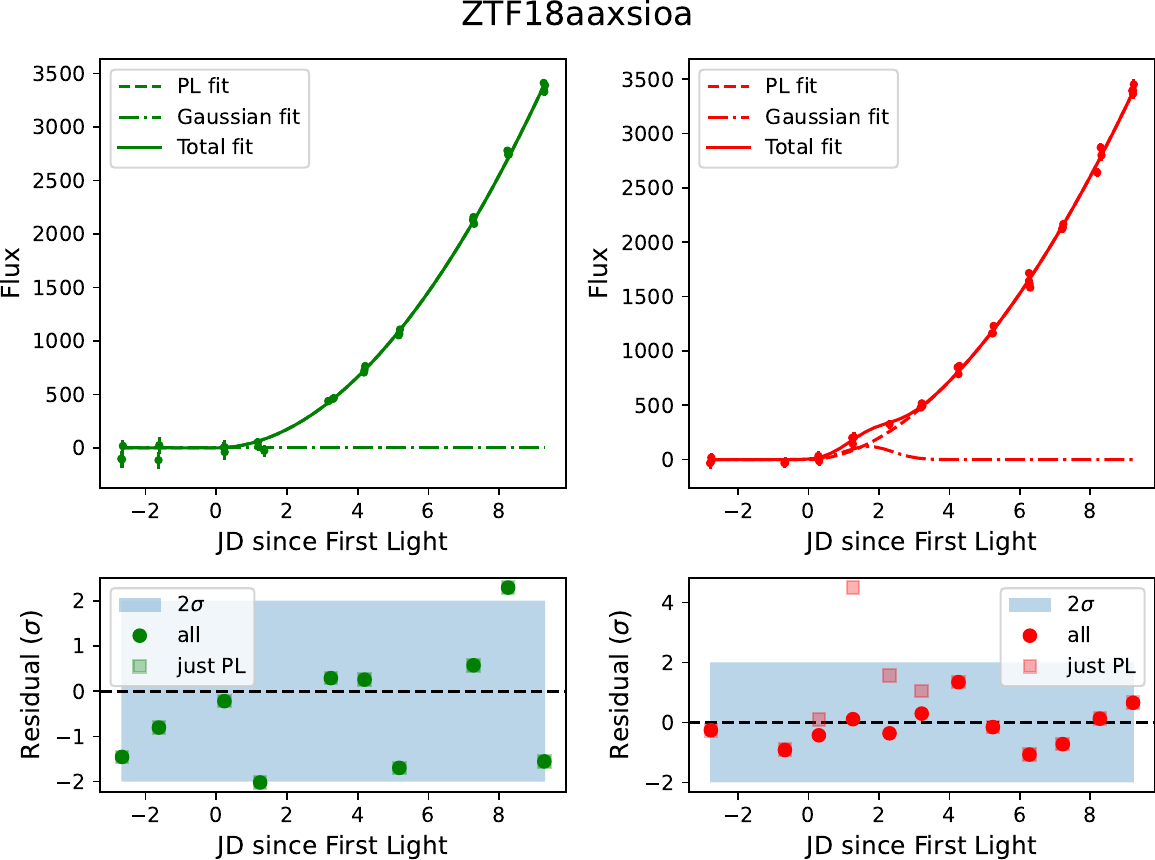}}
         \hfill
    \centering
     \subfloat{
         \centering
         \includegraphics[width=0.48\textwidth]{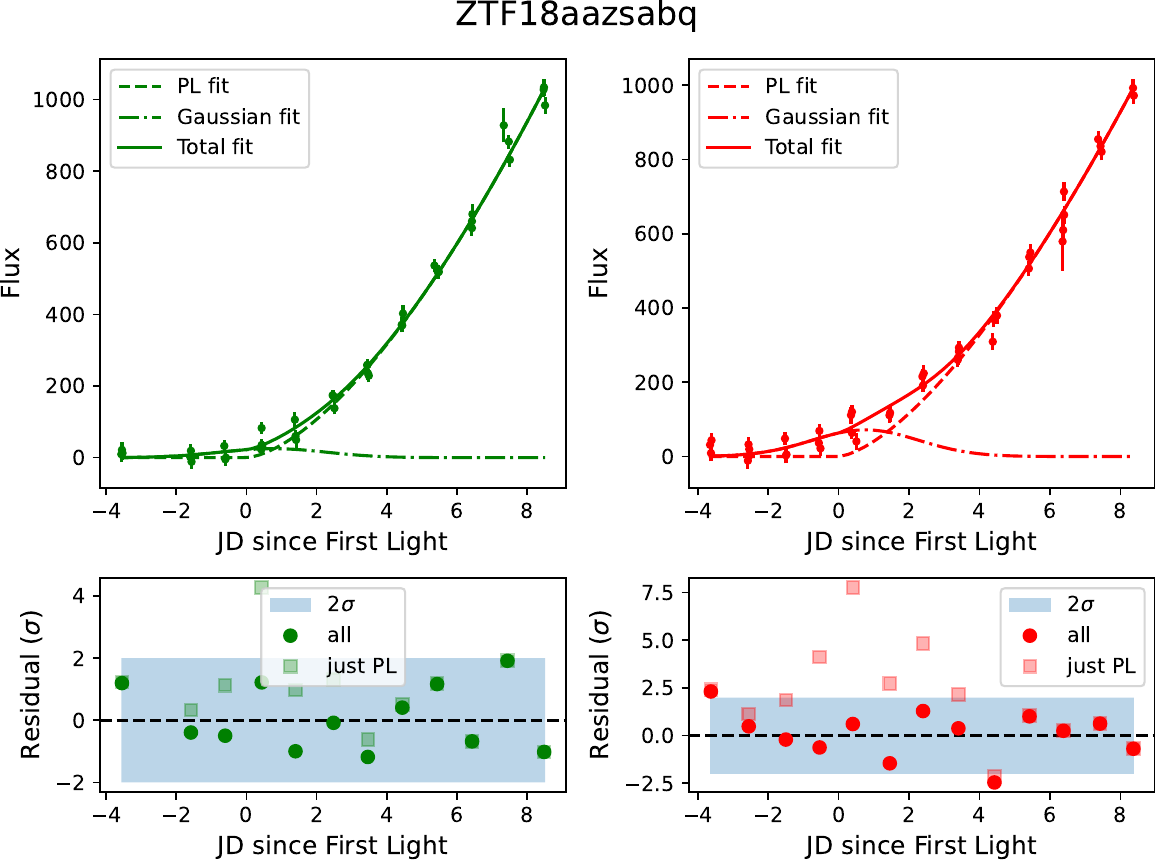}}
         \hfill
     \hfill
     
  \caption{Other early-excess SNe\,Ia from \citet{Yao19}.}
  \label{fig:cont1}
\end{figure*}

\begin{figure*}
     \centering
     \subfloat{
         \centering
         \includegraphics[width=0.48\textwidth]{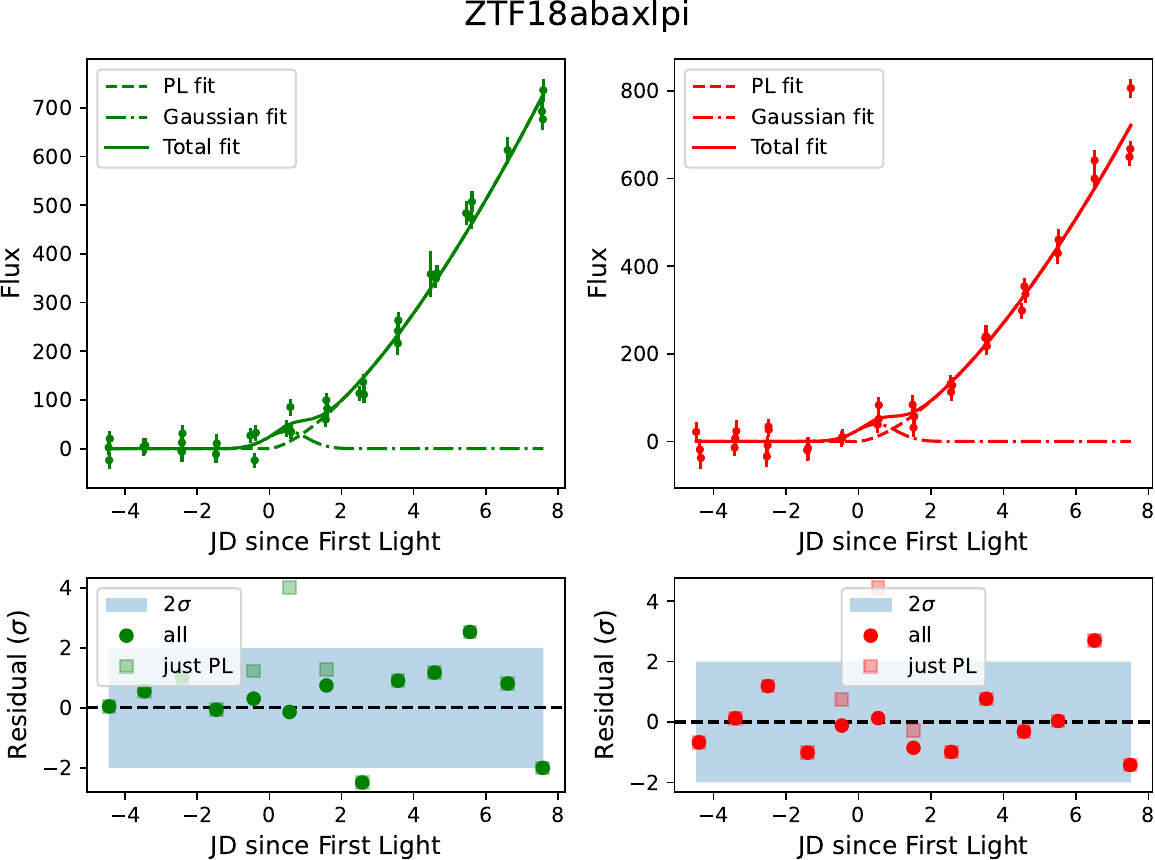}}
         \hfill
    \centering
     \subfloat{
         \centering
         \includegraphics[width=0.48\textwidth]{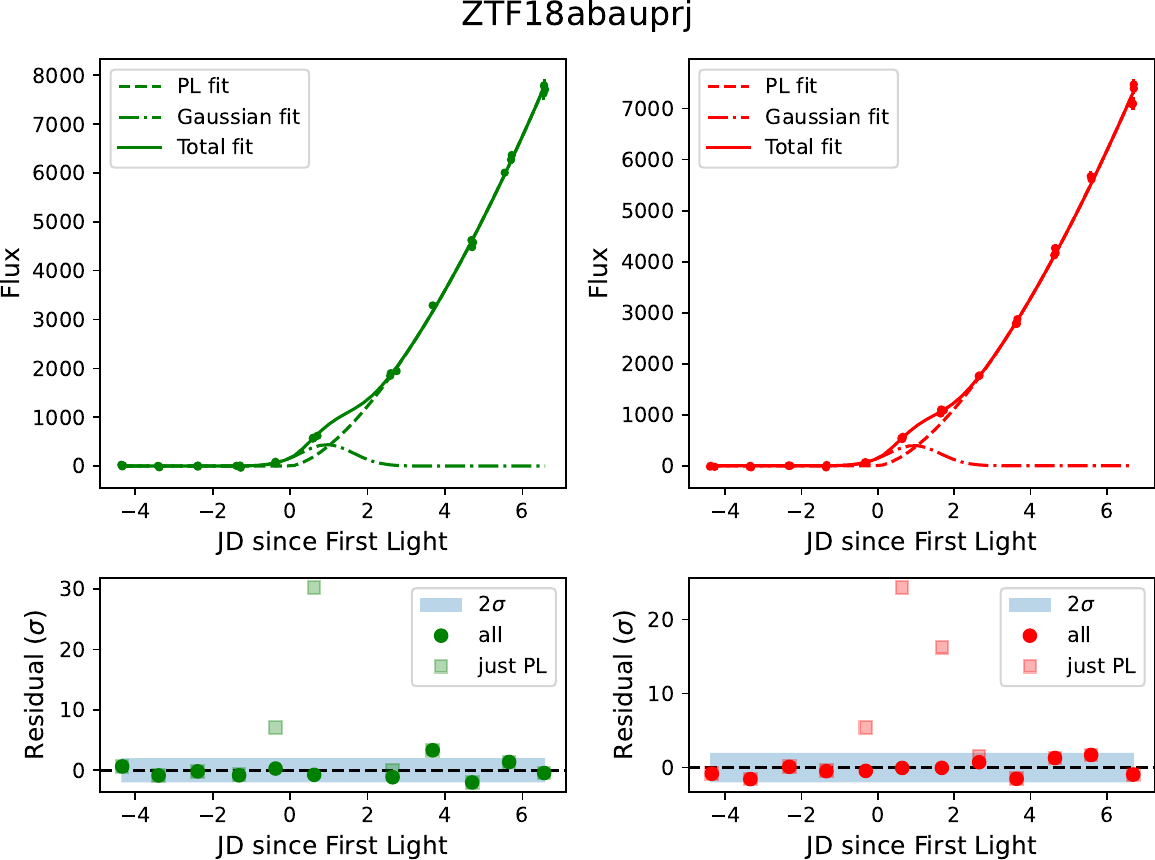}}
         \hfill
     \hfill
     \centering
     \subfloat{
         \centering
         \includegraphics[width=0.48\textwidth]{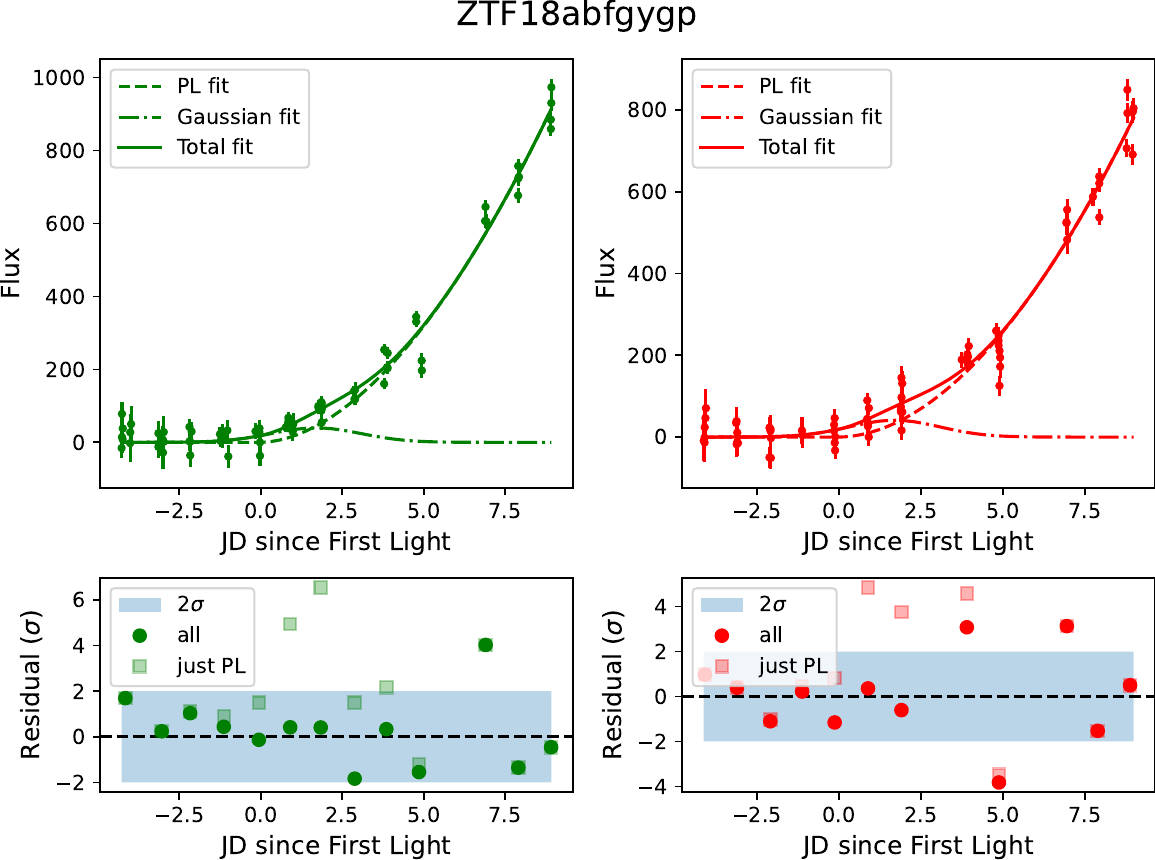}}
         \hfill
    \centering
     \subfloat{
         \centering
         \includegraphics[width=0.48\textwidth]{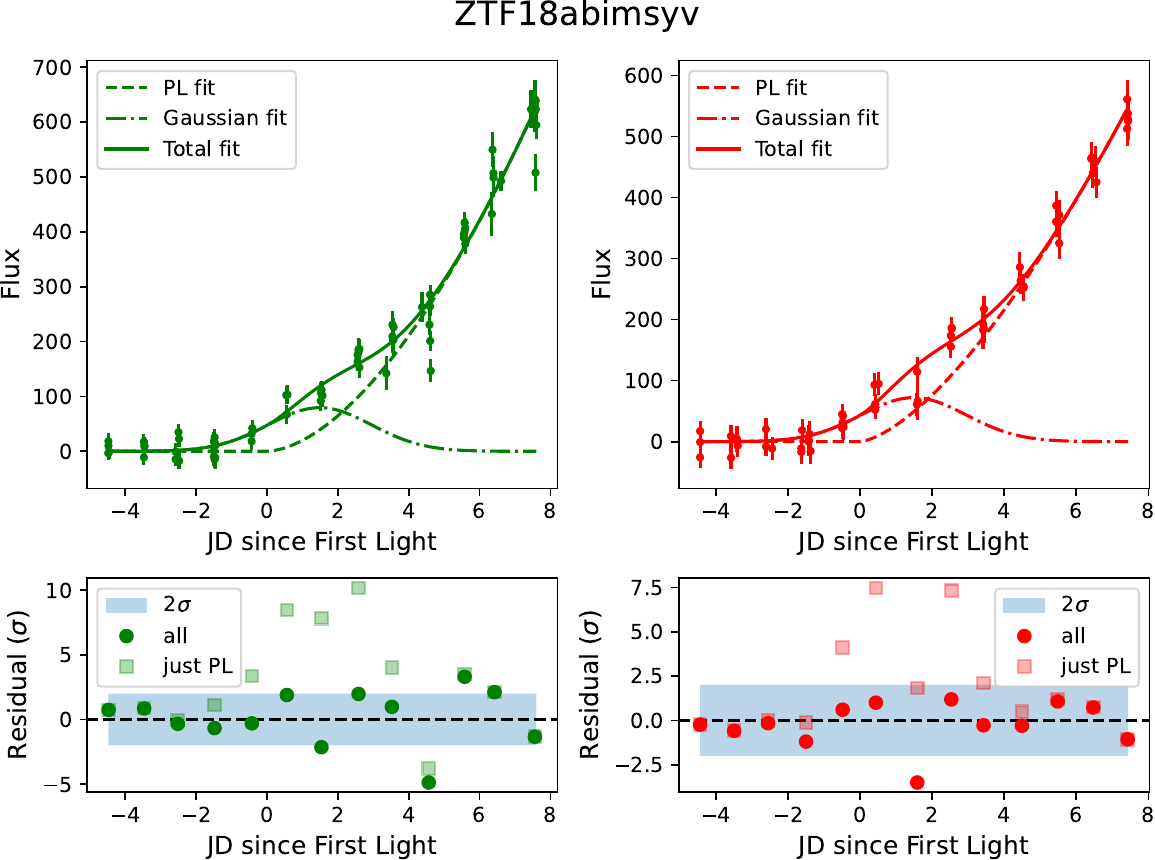}}
         \hfill
     \hfill
     \subfloat{
         \centering
         \includegraphics[width=0.48\textwidth]{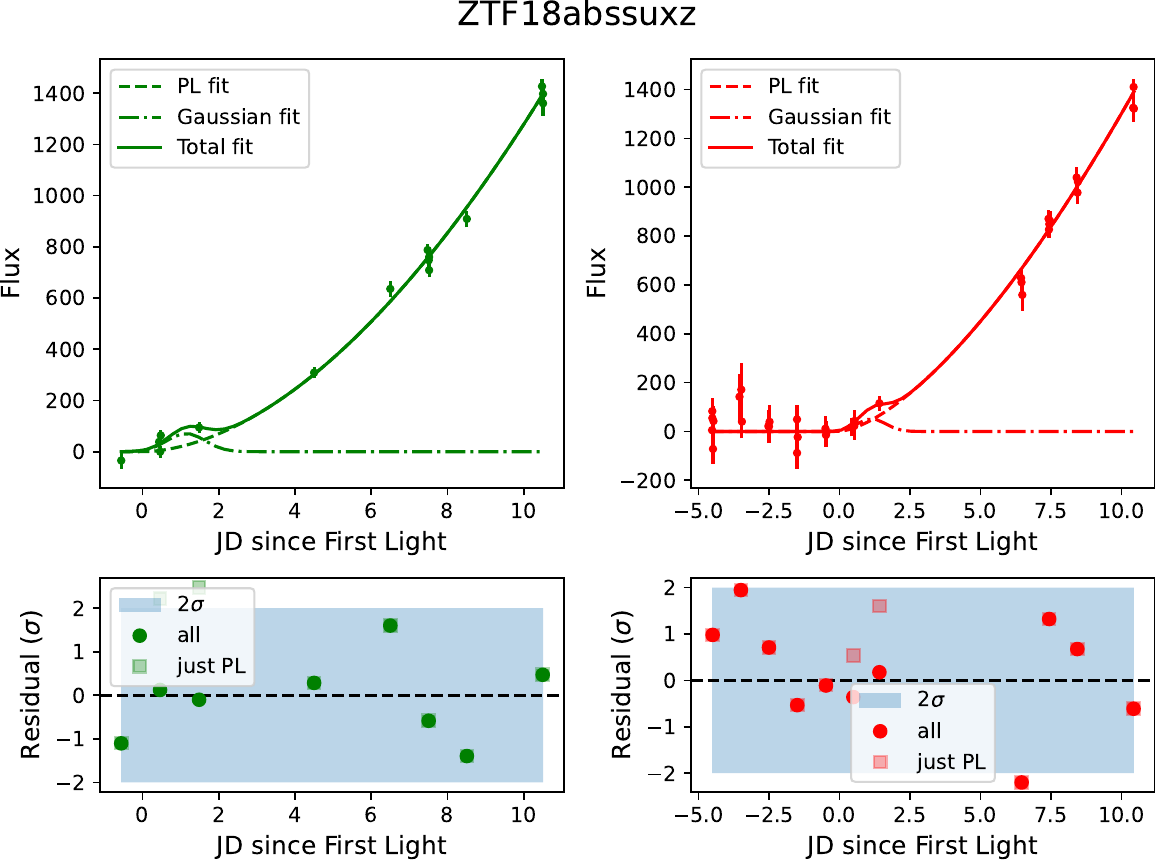}}
         \hfill
    \centering
     \subfloat{
         \centering
         \includegraphics[width=0.48\textwidth]{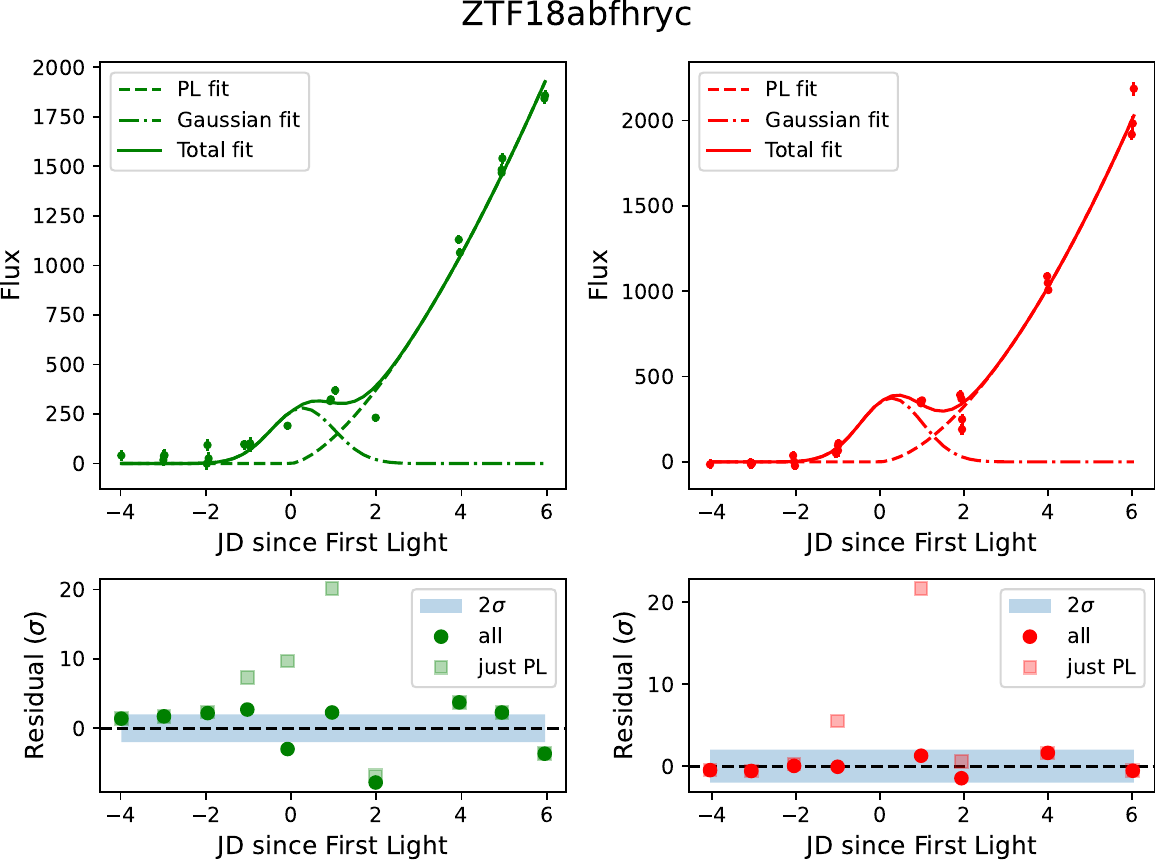}}
         \hfill
     \hfill
  \caption{Other early-excess SNe\,Ia from \citet{Yao19}, contd.}
  \label{fig:cont2}
\end{figure*}

Here, we list the specific criteria for our ``gold" and ``bronze" detections of early excesses, as well as our ``gold" non-detections.  An overview of how these samples are used in our analysis is given in Section \ref{sec:tiers}.  In addition to the criteria below, each light curve has been inspected visually as an additional check on the fidelity of the model fit and data quality.  Our fits to ``gold" and ``bronze" light curves are shown in Figures \ref{fig:detection}, \ref{fig:cont1} and \ref{fig:cont2}.

The gold excess-having sample is defined as follows:

\begin{enumerate}
    \item The SN\,Ia has at least 2 data points between $-5$ and 0 days prior to $t_{exp}$.
    \item For at least 2 consecutive values of $N$ (the maximum epoch used for PL fitting), the BIC for the Gaussian$+$PL model is at least $5$ less than the BIC for the PL-only model.
    \item At least two consecutive data points in a single band favor the Gaussian$+$PL model over the PL-only model at $>3\sigma$ significance.
    \item The reduced $\chi^2$ of the fit is $< 3$.
    \item Data for at least 2 epochs exist within $2\sigma$ of the best-measured excess (i.e., in $[\mu - 2\sigma, \mu + 2\sigma]$).
    \item There are no $>10\sigma$ outliers in the full model fit.
    \item The amplitude of the Gaussian component has at least $3\sigma$ significance and there are no nearby outliers of comparable significance to the detected Gaussian component.
    \item In at least one band, the bump amplitude is at least $2\%$ of the 10-day post-explosion flux.
    \item An excess is still detected even if a stricter ($[1.75, 2.25]$) prior on the PL slope $\alpha$ is enforced.
\end{enumerate}

The bronze excess-having sample is defined similarly, with relaxed criteria:

\begin{enumerate}
    \item The SN\,Ia has at least one pre-explosion data point.
    \item Either at least 2 data points are $>3\sigma$ different between the full Gaussian$+$PL fit and the fit with Gaussian subtracted {\it or} the BIC prefers the Gaussian$+$PL model and at least one data point is $>3\sigma$ different between the Gaussian$+$PL fit and the Gaussian-subtracted fit.
    \item The reduced $\chi^2$ of the fit is $< 6$.
    \item Data for at least one epoch exists within $2\sigma$ of the best-measured excess (i.e., in $[\mu - 2\sigma, \mu + 2\sigma]$).
    \item The SN\,Ia has $\leq 1$ extreme, $>10\sigma$ outliers in the full model fit.
    \item The ``bump" amplitude is at least 1\% of the 10-day post-explosion flux in at least one band.
    \item There are no nearby outliers of comparable significance to the detected Gaussian component.
\end{enumerate}

Finally, we define a gold tier for no-excess SNe:

\begin{enumerate}
    \item Has at least 2 data points pre-explosion (between $-5$ and 0 days).
    
    \item The BIC for the Gaussian$+$PL model is greater than the BIC for the PL-only model.
    
    \item There are no significant (3$\sigma$) differences in any data point between the Gaussian$+$PL model and the same model after the Gaussian component has been subtracted.
    
    \item The reduced $\chi^2$ of the fit is $< 3$.
    
    \item The SN\,Ia has at least 2 data points between $t_{exp}$ and $t_{exp}+5$ (the time range when we would expect an early excess to occur) in each band.
    
    
    \item There are no extreme ($>10\sigma$) outliers in the full model fit.
    
    \item The amplitude of the Gaussian component is $<2\sigma$ significant and less than $2\%$ of the 10-day post-explosion flux in both bands.

    \item There are no claims of an early excess in the literature \citep[e.g.,][]{Yao19,Bulla20,Miller20,Deckers22,Burke22b}.
    
\end{enumerate}

\bibliography{main}

\end{document}